\documentstyle[prd,aps,preprint,tighten,epsfig]{revtex}
\begin{document}
\preprint{                                                 BARI-TH/365-99}
\draft
\title{ 	Three-flavor MSW solutions of the solar neutrino problem}
%
\author{
			G.\ L.\ Fogli~$^a$, 
			E.\ Lisi~$^a$, 
			D.\ Montanino~$^b$, and 
			A.\ Palazzo~$^a$
}
\address{     	$^a$~Dipartimento di Fisica and Sezione INFN di Bari,\\
                   	Via Amendola 173, I-70126 Bari, Italy}
\address{	$^b$~Dipartimento di Scienza dei Materiali 
			dell'Universit\`a di Lecce,\\
             Via Arnesano, Collegio Fiorini, I-73100 Lecce, Italy}
\maketitle
\begin{abstract}
We perform an updated phenomenological analysis of the 
Mikheyev-Smirnov-Wolfenstein (MSW) solutions of the solar neutrino problem,
assuming oscillations between two and three neutrino families. The analysis
includes the  total rates of the Homestake, SAGE, GALLEX, Kamiokande and 
Super-Kamiokande experiments, as well as  the day-night asymmetry and the
18-bin energy spectrum of Super-Kamiokande. Solutions are found at several
values of the $\theta_{13}$ mixing angle. Among the most interesting features,
we find that solar neutrino data alone put the constraint $\theta_{13}\lesssim
55^\circ$--$59^\circ$ at 95\% C.L.,  and that a fraction of the MSW solutions
extends at and beyond maximal $(\nu_1,\nu_2)$ mixing $(\theta_{12}\geq
\pi/4)$,  especially if the neutrino square mass splitting is in its lower
range  $(m^2_2-m^2_1\sim 10^{-7}$ eV$^2$) and if $\theta_{13}$ is nonzero. In
particular, bimaximal (or nearly bimaximal) mixing is possible  for atmospheric
and MSW solar neutrino oscillations within the stringent reactor bounds on
$\theta_{13}$. 
\end{abstract}

\section{Introduction}
\label{s1}

It is widely recognized that the combined sources of evidence for neutrino
flavor transitions coming from the solar neutrino problem \cite{NuAs} and  from
the atmospheric neutrino anomaly\cite{To99}  demand an approach in terms of
three-flavor oscillations among massive neutrinos $(\nu_1,\nu_2,\nu_3)$
\cite{Olds}. The three-flavor $\nu$ parameter space is then spanned by six
variables:    
\begin{mathletters}
\begin{eqnarray}
\delta m^2 	&=& m^2_2 - m^2_1		\ ,\\
 m^2 		&=& m^2_3 - m^2_2		\ ,\\
\omega 		&=& \theta_{12}\in [0,\pi/2]	\ ,\\
\phi 		&=& \theta_{13}\in [0,\pi/2]	\ ,\\
\psi 		&=& \theta_{23}\in [0,\pi/2]	\ ,\\
\delta 		&=& {\rm CP\ violation\ phase}  \ ,
\end{eqnarray}
\label{e1}
\end{mathletters}
where the $\theta_{ij}$ rotations are conventionally ordered as for the quark
mixing matrix \cite{KuoP}.

In the phenomenologically  interesting limit $|\delta m^2|\ll |m^2|$, the two
eigenstates closest in mass $(\nu_1,\nu_2)$ drive solar neutrino oscillations,
while the ``lone'' eigenstate $\nu_3$ drives atmospheric neutrino oscillations.
In such a limit (see \cite{Olds,KuoP}  and refs.\ therein): $i)$ the phase
$\delta$ becomes unobservable; $ii)$ the atmospheric parameter space is spanned
by $(m^2,\psi,\phi)$; and $iii)$ the solar neutrino parameter space is spanned
by $(\delta m^2,\omega,\phi)$.%
\footnote{ In the special case $\phi=0$,  the atmospheric and solar parameter
spaces are decoupled into the two-family oscillation spaces $(\delta
m^2,\omega)$ and $(m^2,\psi)$.}

Building on a previous work \cite{3MSW}, we perform a thorough analysis of the
available solar neutrino data in the $(\delta m^2,\omega,\phi)$ variables, in
the context of the Mikheyev-Smirnov-Wolfenstein (MSW) oscillation mechanism
\cite{MSWm}.  There are several motivations to revisit, update and improve the
analysis in \cite{3MSW},  the most important being the need to include the
high-statistics Super-Kamiokande (SK) observations of neutrino events (total
rate, \cite{To99}, energy spectrum \cite{To99,SKSP}, and  day-night difference 
\cite{To99,SKDN}). Indeed, the few post-SK papers on three-flavor  MSW
oscillations we are aware of \cite{St98,Te99,Ki99}  include neither the
spectrum information nor day-night variations. This contrasts with the more
familiar two-family MSW analyses, which have been regularly updated   with
state-of-the-art global fits at different  detector lifetime days; see, e.g.,
the SK official fits (\cite{To99,Su99,Na99} and refs.\ therein),  as well as
various analyses by independent research groups: \cite{100d} (100 days),
\cite{300d} (300 days), \cite{500d} (500 days), and \cite{800d,SNOL} (800
days).

After the work \cite{3MSW}, there have been also other relevant experimental
and theoretical improvements (implemented in the present analysis): updated
measurements  of the experimental (total) rates in the chlorine (Cl)
\cite{Cl98} and Kamiokande (K) \cite{Fu96} experiments and in the gallium (Ga)
detectors SAGE  \cite{Ab99} and GALLEX \cite{Ha99}; new standard solar model
(SSM) estimates for the neutrino fluxes and their uncertainties, and for the
neutrino production regions in the Sun \cite{BP98}; updated calculations of the
$^8$B neutrino spectrum \cite{Al96}, of  the $\nu_{e,\mu}$ scattering cross
section on electrons \cite{Ka95}, and of the $\nu_e$ neutrino absorption cross
section in chlorine \cite{Al96} and in gallium \cite{GaCs}; analytical methods
to compute the (time-averaged) $\nu_e$ survival probability in the Earth
\cite{Li97}.

In addition, some relatively old topics,   such as the role played by a
possibly large {\em hep\/} neutrino flux \cite{Kuzm,Rhep}, or by solar neutrino
mixing at and beyond ``maximal'' values  $(\omega\geq \pi/4)$ \cite{3MSW},  are
currently being revisited (see \cite{Heps,Fi98,Ho99} and \cite{Gu99,Mu99,Last},
respectively) and  demand an updated discussion.  In particular, the case of
maximal   solar neutrino mixing $(\omega=\pi/4)$ appears now more ``natural''
and interesting in light of the Super-Kamiokande results on atmospheric
neutrinos, which favor maximal $\nu_\mu\leftrightarrow\nu_\tau$ mixing ($\psi=
\pi/4$).%
\footnote{Solar $\nu$ MSW oscillations at $\sin^2 2\omega=1$  are nontrivial 
when Earth matter effects are included \cite{3MSW,Gu99,Mu99,Last}. This fact
was not considered in Refs.~\protect\cite{Gi98,GiBy,Ha96}, where only  matter
effects in the Sun were studied.}
Therefore, it is important to go beyond the usual analyses in terms of the
solar $\nu$ mixing parameter  $\sin^2 2\omega$ (equivalent to take $\omega\leq
\pi/4$), which miss  a potentially interesting region of the MSW solutions at
$\omega>\pi/4$, as we shall see later (see also \cite{3MSW} for earlier
discussions).

Concerning the parameter $\phi$, we do not include {\em a priori\/} the
constraints coming from the CHOOZ reactor experiment \cite{CHOO}, which,
together with the atmospheric neutrino  data \cite{To99}, imply that
$\sin^2\phi \lesssim $ few \% \cite{3atm,Last} (the exact upper limit depending
on the value of $m^2$, on  the confidence level chosen, and on the number of
degrees of freedom in the oscillation model).  In fact, we think that it is
instructive to study the constraints  on $\phi$ coming from solar neutrino data
{\em alone\/}, as it has been done similarly for atmospheric neutrino data 
\cite{3atm}.

In any case, observable effects  on the MSW solutions can be generated even by
small (few \%) values of $\sin^2\phi$, which induce a fractional suppression in
the solar neutrino survival probability $P_{ee}$ approximately equal to
$2\times \sin^2\phi$. In fact, if $N_e$ is the electron density profile, the
$3\nu$ and $2\nu$ expressions for $P_{ee}$  are related by the expression
\begin{equation}
P_{ee}^{3\nu}(\delta m^2,\omega,\phi) = \sin^4\phi +
\cos^4\phi\cdot P_{ee}^{2\nu}(\delta m^2,\omega)
\Big|_{N_e\to \cos^2\phi \,N_e}\ ,
\label{e2}
\end{equation}
which, for small values of $\phi$, gives roughly 
\begin{equation}
\frac{P_{ee}^{3\nu} - P_{ee}^{2\nu}}{P_{ee}^{2\nu}}\simeq  -2\,\sin^2 \phi\ .
\label{e3}
\end{equation}
Even within the stringent CHOOZ limits on $\sin^2 \phi$, such variations of
$P_{ee}$ can become as large as the uncertainties on the neutrino event rates, 
and thus cannot be neglected.

The above motivations warrant the present  work, which is organized as follows.
In Sec.~\ref{s2}  we present the theoretical and experimental  ingredients of
the analysis. In Sec.~\ref{s3} we update the familiar two-family oscillation
fit ($\phi=0$), and discuss in detail the features of the $2\nu$ MSW solutions.
In Sec.~\ref{s4} we extend the analysis to three-flavor oscillations, and
discuss how the $3\nu$ MSW solutions change for increasing values of $\phi$. We
summarize our results in Sec.~\ref{s5}. The details of the statistical analysis
are given in the Appendix.

\section{Standard predictions and experimental data}
\label{s2}

In the present analysis, we use the so-called BP98 standard solar model
\cite{BP98} for the electron density in the Sun and for the input neutrino
parameters ($\nu_e$ fluxes, spectra, and production regions), and compare the
predictions to the experimental data for the following observables: total
neutrino event rates,  SK energy spectrum, and SK day-night asymmetry.

Table~\ref{tablerates} shows the latest results (and standard expectations) for
the total neutrino event rates measured  at Homestake \cite{Cl98}, Kamiokande
\cite{Fu96},  SAGE \cite{Ab99}, GALLEX \cite{Ha99}, and Super-Kamiokande (825
live days) \cite{To99,Su99,Na99}. Since the SAGE and GALLEX detectors measure
exactly the same quantity (up to a negligible difference in the  detector
latitude),  their results are combined in a single (Ga) rate of $72.5\pm 5.6$
SNU.  On the other hand, the Kamiokande and Super-Kamiokande data are treated
separately (rather than combined in a single datum), since the two experiments,
although based on the same $\nu$-$e$ scattering detection technique, have
rather different energy  thresholds and resolution functions.%
\footnote{The inclusion of the Kamiokande rate is currently not decisive in
shaping the MSW solutions, the SK rate being much more accurately measured.
However, its addition to the Cl, Ga, and SK rates avoids  a situation of ``zero
degrees of freedom'' in the $\chi^2$ fit to the total rates for the $3\nu$ MSW
case [three data (Cl, Ga, SK) minus three free parameters  $(\delta m^2,
\omega, \phi)$]. In such a situation, the value of  $\chi^2_{\min}$ would not
have a well-defined likelihood.}

The SK electron recoil energy spectrum and its statistical and systematic
uncertainties (825 lifetime day, $E_e>5.5$ MeV) are graphically reduced from
the 18-bin histograms shown by SK members in recent Summer~'99 conferences
\cite{To99,Su99,Na99}. The corresponding numerical values have been already
reported in detail in Ref.~\cite{800d} and are not repeated here. Our
theoretical calculation of the binned spectrum properly takes into account
energy threshold and resolution effects (see, e.g., the Appendix of
Ref.~\cite{Fa99}). Standard $^8$B \cite{Al96} and {\em hep\/} \cite{BP98}
neutrino spectra and fluxes are used, unless otherwise noted.  Concerning the
SK day-night asymmetry of the event rates, we use the latest measurement
\cite{Na99}:
\begin{equation}
2\;\frac{N-D}{N+D}=0.065\pm0.031\pm0.013\ .
\label{e4}
\end{equation}

In the presence of $2\nu$ or $3\nu$ oscillations,   the MSW effect in the Sun
is computed as described in Ref.~\cite{3MSW}. The additional Earth matter
effects are treated as in Ref.~\cite{Li97}. The  $\chi^2$ analysis of the
theoretical and experimental uncertainties basically follows the approach
developed in \cite{Fo95}, with the necessary updates to take into account the
BP98 SSM predictions and  the energy spectrum information. Technical details
about error estimates are given in the Appendix.

We conclude this section by comparing the standard (no oscillation) predictions
with the experimental data for the Cl, Ga, and SK total rates.  Figure~\ref{f1}
shows the 99\% C.L.\ error ellipses for data and expectations in the planes
charted by the (Cl,~Ga), (SK,~Ga), and (SK,~Cl) total rates. The  distance
between observations and  standard predictions makes the solar neutrino
problem(s) evident. At present, such information is the  main evidence for
solar neutrino physics beyond the standard electroweak model; however, since
the theoretical errors are dominant---as far as total rates are
concerned---no  substantial improvements can be expected by a reduction of the
experimental errors. Conversely, decisive information is expected  from the SK
spectrum and day-night asymmetry, but no convincing deviation has emerged from
such data yet. Therefore, it is not surprising that, in oscillation fits, the
total rates mainly determine {\em allowed\/} regions, while the SK spectrum and
day-night asymmetry determine {\em excluded} regions.

\section{Two-flavor MSW oscillations}
\label{s3}

Figure~\ref{f2} shows the results of our $2\nu$ MSW analysis of the data
discussed in the previous section, shown as confidence level contours in the
$(\delta m^2,\sin^2 2\omega/\cos 2\omega)$ plane. The choice of  the variable
$\sin^2 2\omega/\cos 2\omega$, rather than the usual $\sin^2 2\omega$, allows
an expanded view of the large mixing region.  In each of the six panels, we
determine the absolute minimum of the $\chi^2$ and then plot the iso-$\chi^2$
contours at $\chi^2-\chi^2_{\min}=4.61$, 5.99, and 9.21, corresponding to
90\%, 95\%, and 99\% C.L.\ for two degrees of freedom (the oscillation
parameters). In fits including the total rates,  there is a global  $\chi^2$
minimum  and two local mimima; such minima, and the surrounding favored
regions, are usually indicated as MSW solutions at small mixing angle (SMA),
large mixing angle (LMA), and low $\delta m^2$ (LOW). Additional information on
such solutions is reported in  Tab.~\ref{chisquare}.

Concerning the statistical interpretation of the $\chi^2$ values, a remark is
in order. One can attach confidence levels to $\chi^2$ values in two different
ways, depending on the choice between  {\em hypotheses tests\/} and {\em
parameter estimation\/} \cite{Stat}.  If one is interested in testing the
goodness of the MSW hypothesis {\em a priori}, then one should compare the
absolute $\chi^2_{\min}$ with a number of degrees of freedom $(N_{\rm DF})$
calculated as number of data minus number of free parameters. The corresponding
probability $P$ is given in the last column of Tab.~\ref{chisquare} where, for
completeness, $P$ is reported also for the other two local minima and for the
no oscillation case. If the MSW hypothesis is accepted, then the MSW parameter
estimation involves only $\chi^2$ differences with respect to the global
minimum, and the appropriate value of $N_{\rm DF}$ to use is the number of free
parameters in the model ($N_{\rm DF}=2$), as  anticipated for Fig.~\ref{f2}.

The first panel of Fig.~\ref{f2} refers to the fit to the total rates only. The
three $\chi^2$ minima are indicated by dots. The absolute minimum is reached
within the SMA solution $(\chi^2_{\min}=1.08)$, which represents a very good
fit to the data. The LMA solution is also acceptable, while the LOW solution
gives a marginal fit (see also the upper three rows of Tab.~\ref{chisquare}).
The SK data on the day-night asymmetry (second panel) and  energy spectrum
(third panel) exclude large regions in the mass-mixing parameter space; but are
unable to (dis)prove any of the three solutions, which in fact are present also
in the global fit to all data (fifth panel; see also the middle three rows of
Tab.~\ref{chisquare}).

The spectrum information is somewhat sensitive to the (uncertain) value of the
{\em hep\/} neutrino flux; for instance, an enhancement by a factor $20$ helps
to fit the high-energy part of the SK spectrum \cite{Heps}, and thus it
produces a reduction of the excluded regions  in the mass-mixing plane  (fourth
panel in Fig.~\ref{f2}), and a corresponding slight enlargement of the globally
allowed regions (sixth panel; see also the lower three  rows of
Tab.~\ref{chisquare}).%
\footnote{The SK Collaboration \protect\cite{Su99} derives a preliminary upper
limit of $\sim 15$ to the {\em hep}/SSM flux ratio (in the absence of
oscillations). This limit is weakened to $\lesssim30$ in the presence of an
oscillation effect of $\sim 50\%$.  Therefore, we  can assume $20\times{\em
hep}$ as a relevant, representative case of large {\em hep\/} flux.}

From the results of Fig.~\ref{f2} and of Tab.~\ref{chisquare}, it appears that
the inclusion of the day-night and spectral information can significantly
change the C.L.'s associated to each of three solutions SMA, LMA, and LOW. In
order to understand better the role of difference pieces of data, we show in
Fig.~\ref{f3} the comparison between data and predictions for the
$\chi^2_{\min}$ point of the total rate fit (SMA solution at best fit, first
row of Tab.~\ref{chisquare}),  in the same coordinates as in Fig.~\ref{f1}.
Analogously, Figs.~\ref{f4} and  \ref{f5} show the analogous comparison for the
LMA and LOW solutions (second and third row of Tab.~\ref{chisquare}).
Figure~\ref{f6} shows the spectral information for the three local minima in
the global fit to all data with standard {\em hep\/} flux (upper panel) and
with enhanced ($20\times$) {\em hep} flux (lower panel).

From Figs.~\ref{f3}--\ref{f5} it appears (as far as total rates are concerned)
that the SMA solution represents a very good fit to the SK, Ga, and Cl data,
that the LMA solution underestimates slightly the Ga and SK data, and that the
LOW solution underestimates the Ga rate and overestimates the Cl rate. This
explains the ordering in the likelihood of such solutions in
Tab.~\ref{chisquare}:  $P({\rm SMA})>P({\rm LMA})>P({\rm LOW})$. On the other
hand, Figure~\ref{f6} shows that the bulk of the SK observed spectrum is
basically consistent with a flat shape (except for the two highest-energy, 
low-statistics bins), and therefore  tends to favor the LMA and LOW solutions,
rather than the steadily increasing spectrum predicted by the SMA solution.
This tendency is slightly enhanced in the case of large {\em hep\/} neutrino
flux (lower panel of Fig.~\ref{f6}) since in this case the LMA and LOW
solutions can fit somewhat better the spectrum endpoint. In other words, the
inclusion of the spectrum in the global analysis tends to compensate the
different likelihoods of the LMA and LOW solutions with  respect to the SMA
solution. Moreover, the slight excess of  observed nighttime events
[Eq.~(\ref{e4})] adds extra likelihood to the LMA solution, so that in the
global fit one has $P({\rm SMA})\sim P({\rm LMA})>P({\rm LOW})$ (see
Tab.~\ref{chisquare}), with the LMA solution even more likely than the SMA
solution in the case of large {\em hep\/} flux.

From the previous discussion, the following situation emerges for the three MSW
solutions SMA, LMA, and LOW. None of them can be excluded at 99\% C.L.\ by the
present experimental data. Different pieces of data give indications that are
not as consistent as it would be desirable: the total rate information favors
the SMA solution, the spectral data favor the LMA and LOW solutions, and the
day-night data favor the LMA solution. In a global fits, the three solutions
have comparable likelihoods. Although such solutions are subject to change
shape and likelihood as more accurate experimental data become available, no
dramatic improvement    can be really expected in their selection, unless:  1)
the theoretical uncertainties on the total rates are reduced to the size of the
corresponding experimental uncertainties; 2)  the total errors associated to
the SK spectrum and day-night measurement are significantly reduced (by, say, a
factor $\sim 2$); or 3) decisive results are found in new solar neutrino
experiments such as the Sudbury Neutrino Observatory (SNO)
\cite{SNOL,SNim,Vill}, the Gallium Neutrino Observatory \cite{GNOE,Mo99},
BOREXINO  \cite{BORE,Mo99}, and KamLand \cite{KamL}.  Any of these conditions
require a time scale of a few years at least; the same time scale should then
be expected in order to (patiently) single out one of the three MSW solutions
(SMA, LMA, or LOW).

Another aspect of the LMA and LOW solutions emerging from Fig.~\ref{f2} is
their extension to large values of the mixing angle ($\sin^2 2\omega\to 1$),
which are often assumed to be  realized only through the vacuum oscillation
solutions.  Since the  possibility of nearly maximal $(\nu_1,\nu_2)$ mixing 
for solar neutrinos has gained momentum after the SK evidence for maximal
$(\nu_\mu,\nu_\tau)$ mixing $(\sin^2 2\psi\sim 1)$, it is interesting to study
it in detail  by dropping the usual ``$2\omega$'' variable and by exploring the
full range $\omega\in [0,\pi/2]$, as it was done earlier in \cite{3MSW}. The
subcase $\omega=\pi/4$ will receive special attention in  the next section.

\section{Three-flavor MSW oscillations}
\label{s4}

As stated in Sec.~\ref{s1}, for large values of $m^2$ ($\gg 10^{-4}$~eV$^2$)
the  parameter space relevant for $3\nu$ solar neutrino oscillations is spanned
by the variables  $(\delta m^2,\omega,\phi)$. As far as $\omega$ is taken in
its full range $[0,\pi/2]$, one can assume $\delta m^2>0$, since the MSW
physics is invariant under the substitution $(\delta m^2,\omega) \to (-\delta
m^2,\pi/2-\omega)$ at any $\phi$.%
\footnote{Only for solar neutrino oscillations in {\em vacuum\/}, the further
symmetry $\omega\to\pi/2-\omega$ (at any $\delta m^2$  and $\phi\in[0,\pi/2]$)
allows to take $\delta m^2>0$ with $\omega$ in the restricted range
$[0,\pi/4]$.}

For  graphical representations, we prefer to use the mixing variables $(\tan^2
\omega,\tan^2 \phi)$ introduced in \cite{3MSW}, which  properly chart both
cases of small and large mixing. The case $\tan^2\phi=0$ corresponds to the
familiar two-family oscillation scenario, except that now we consider also the
usually neglected case $\omega>\pi/4$ ($\tan^2\omega>1$).  For each set of
observables (rates, spectrum, day-night difference, and  combined data) we
compute the corresponding MSW predictions and their uncertainties, identify the
absolute minimum of the $\chi^2$ function, and determine the surfaces at
$\chi^2-\chi^2_{\min}=6.25$, 7.82, and 11.36, which   define the volumes
constraining the $(\delta m^2,\tan^2\omega,\tan^2\phi)$ parameter space at
90\%, 95\%, and 99\% C.L. Such volumes are graphically presented in $(\delta
m^2,\tan^2 \omega)$ slices for representative values of $\tan^2\phi$.

Figure~\ref{f7} shows the results for the fit to the total rates only.  The
absolute minimum $(\chi^2_{\min}=0.7)$ is reached within the SMA solution at
nonzero $\tan^2\phi$, as reported in the first row of Tab.~\ref{chiminima}. The
preference for $\phi\neq 0$ is, however,  not statistically significant, since
$\chi^2_{\min}(\phi=0)-\chi^2_{\min}=1.08-0.70=0.38$ only. The qualitative
behavior of the $3\nu$ MSW solutions in Fig.~\ref{f7} is unchanged with respect
to the earlier analysis in \cite{3MSW}: for increasing  $\tan^2\phi$, the LOW
solution moves to slightly larger values of $\tan^2\omega$ and then disappears,
while the SMA and LMA regions  tend to merge in a single, broad solution which
slowly disappears at large values of $\phi$. In general, the MSW allowed
region(s) become less structured for increasing $\phi$, due to the the
decreasing energy dependence of the $\nu_e$ survival probability \cite{3MSW}. 
In the first panel of Fig.~\ref{f7} it is interesting to note that both the LMA
and the LOW solutions are compatible at 99\% C.L.\ with maximal mixing
$(\tan^2\omega=1)$. This compatibility persists for the LOW solution at small
values of $\tan^2\phi$, while it is lost for the LMA solution, which migrates
towards smaller mixing angles.  As for the two-family case discussed in the
previous section, the addition of the spectrum information (and, to a lesser
extent, of the day-night asymmetry) is expected to produce significant changes
in the C.L.'s associated to the MSW solutions in Fig.~\ref{f7}.

Figure~\ref{f8} shows the region excluded by the SK day-night asymmetry, in the
same representation as in Fig.~\ref{f7}. Such region is generally far from the
allowed regions in Fig.~\ref{f7}, and thus the day-night asymmetry contributes
only marginally to shape the MSW solutions, except at low values of
$\tan^2\phi$, where it  cuts the lower part of the LMA solution and  disfavors
the rightmost part of the SMA solution. The region excluded by the SK day-night
asymmetry can extend beyond maximal mixing  $(\tan^2\omega >1)$, as it was
observed earlier in Ref.~\cite{3MSW} for the corresponding Kamiokande variable.

Figure~\ref{f9} shows instead the regions excluded by the SK spectrum, which
plays a relevant role in disfavoring the zone where the SMA and LMA solutions
tend to merge (see the six middle panels of Fig.~\ref{f7}). In fact, for
$\delta m^2\sim 10^{-4}$ eV$^2$ the normalized SK spectrum is predicted to have
a large {\em negative\/} slope (see, e.g., \cite{Li97}), contrary to the
experimental results. Relatively large positive values for the spectrum slope
are excluded in the left diagonal band shown in the upper panels of
Fig.~\ref{f9}. Both positive and negative values for the spectrum slope can
instead occur in the excluded region where the Earth regeneration effects are
relevant \cite{Li97}. For increasing $\phi$, the spectral information becomes
less effective in disfavoring zones of the parameter space, since the
energy-dependence of the electron survival probability becomes weaker.
Therefore, only the strongest distortion effect  (at $\delta m^2 \sim 10^{-4}$
eV$^2$) survives at large $\tan^2\phi$. Finally, as in the $2\nu$ case, the
excluded regions are  slightly shrinked if large {\em hep\/} flux values are
assumed (not shown).

Figure~\ref{f10} shows the combined fit to all data, which is one one of the
main results of this work. As for the total rates, the minimum $\chi^2$ (second
row of Tab.~\ref{chiminima}) is reached within the SMA solution and shows a
very weak preference for nonzero values of $\phi$ ($\tan^2\phi\simeq 0.1$).  By
comparing Fig.~\ref{f10} with Fig.~\ref{f7}, it can be seen that the SK
spectrum  excludes a significant fraction of the solutions at  $\delta m^2\sim
10^{-4}$ eV$^2$, including the upper part of the LMA solution at small $\phi$,
and the merging with the SMA solution at large $\phi$. In particular, at
$\tan^2\phi=0.1$  the 95\% C.L.\ upper limit  on $\delta m^2$ drops from 
$2\times 10^{-4}$ eV$^2$ (rates only) to $8\times 10^{-5}$ eV$^2$  (all data).
This indication tends to disfavor neutrino searches of {\sl CP\/} violation
effects, since such effects decrease with $\delta m^2/m^2$ at nonzero values of
$\phi$.

Figure~\ref{f10} also shows that the inclusion of the present spectrum and
day-night information, rather than helping in constraining $\tan^2\phi$, tends
to slightly weaken its upper bound: for instance, the region allowed at 95\%
C.L.\ in the last panel of Fig.~\ref{f10} is larger than the corresponding one
in Fig.~\ref{f7}. In fact, the information coming from total rates prefers some
energy dependence in the survival probability $P_{ee}$, and thus disfavors
large values of $\phi$  [see Eq.~(\ref{e2})]. However, since no clear
indication  for $\partial P_{ee}/\partial E_\nu\neq 0$ emerges from the bulk of
the SK spectrum, its addition in the fit  makes large values of $\phi$
slightly  more ``acceptable.'' In addition, at $\tan^2\phi \gtrsim 1$, the
lower part  of the broad region allowed by total rates (last three panels of
Fig.~\ref{f7}) becomes more consistent with the (weak) indication for a nonzero
night-day difference [Eq.~(\ref{e2})], thus giving some extra allowance for
large values of $\phi$, as shown in the last three panels of Fig.~\ref{f10}.

Figure~\ref{f11} is analogous to Fig.~\ref{f10}, the only difference being  the
flux of {\em hep\/} neutrinos (increased by a factor of 20 with respect to
BP98). Since a large {\em hep} flux allows a better fit to the SK spectrum
endpoint at any value of $\phi$, the global fit is somewhat improved (third row
of Tab.~\ref{chiminima}), and the MSW solutions are slightly more extended than
in Fig.~\ref{f10}. As for the $2\nu$ case, the best-fit point migrates from the
SMA to the LMA solution. In general, the comparison of Figs.~\ref{f10} and 
\ref{f11} shows that the (uncertain) value of the {\em hep\/} neutrino flux is
not crucial in shaping the current $3\nu$ MSW solutions (although it might
become decisive with more accurate spectral data).

Figure~\ref{f12} shows the values of $\chi^2-\chi^2_{\min}$ as a function of
$\tan^2\phi$ for unconstrained $\delta m^2$ and $\omega$, coming from the fit
to total rates and to all data. The values of $\chi^2_{\rm min}$ can be read
from Tab.~\ref{chiminima}. This figure displays more clearly some anticipated
features of the current $3\nu$ MSW solutions (for standard {\em hep\/} flux): 
1) a slight preference for nonzero values of $\tan^2\phi$; 2) the existence of
an upper bound on $\phi$, slightly stronger in the fit to total rates ($\phi
\lesssim 55^\circ$ at 95\% C.L.) than in the global fit $(\phi \lesssim
58^\circ)$. In the case of  enhanced ($20\times$) {\em hep\/} flux we find 
that: 1) the absolute minimum of $\chi^2$ is reached at $\tan^2\phi=0$ within
the LMA solution; 2) for fixed $\phi\neq 0$, the local minimum migrates to the
SMA solution when $\tan^2\phi\gtrsim 0.05$; and 3) the 95\% C.L.\ upper limit
on $\tan^2\phi$ is slightly weakened to $\sim 59^\circ$.

The 95\% C.L.\ upper bound on $\phi$ coming from solar neutrino data alone
$(\phi \lesssim 55^\circ$--$59^\circ)$ is consistent with the one coming from
atmospheric neutrino data alone $(\phi \lesssim 45^\circ)$ \cite{3atm}, as well
as with the upper limit coming from the combination of CHOOZ and atmospheric
data  $(\phi \lesssim 15^\circ)$ \cite{Last,3atm}. This indication supports the
possibility that solar, atmospheric, and CHOOZ data can be interpreted in a
single three-flavor oscillation framework \cite{3MSW,3atm}. In this case, the
CHOOZ constraints on $\phi$ exclude a large part of the $3\nu$ MSW parameter
space (basically all but the first two panels in  Figs.~\ref{f7}--\ref{f10}).

However, even small values  of $\phi$ can be interesting for solar $\nu$
phenomenology, as anticipated in the comment to Eq.~(\ref{e3}).
Figure~\ref{f13} shows, for instance, how the partial and global fits to 
neutrino event rates (Cl, Ga, and SK experiments) are modified when passing
from $\tan^2\phi=0$ to $\tan^2\phi=0.06$. Each of the three experiments, taken
separately, allows $\omega \leq \pi/4$. In the combination,  the LOW and LMA
solutions ``touch'' the $\omega= \pi/4$ line at $\tan\phi=0$, while for nonzero
$\phi$ only the LOW solution remains consistent with $\omega=\pi/4$ (and, to
some extent, with $\omega>\pi/4$). This fraction of the LOW solution, usually
ignored in analyses using the $\sin^2 2\omega$ variable, can be relevant for
model building. In particular, there is currently great interest in models
predicting exact or nearly bimaximal mixing, namely, 
$(\omega,\psi,\phi)\simeq(\pi/4,\pi/4,0)$ (see, e.g., \cite{Bmax,Smir} and
refs.\ therein). Such models imply, for the neutrino mixing matrix $U_{\alpha
i}$, that $U^2_{e1}\simeq U^2_{e2}$ and that $U^2_{e3}\simeq 0$. Recent
theoretical predictions for (small) nonzero values of $U^2_{e3}$ are 
discussed, e.g., in \cite{Ahkm}.

Figure~\ref{f14} shows the section of the volume allowed in the $3\nu$ MSW
parameter space, for $\omega=\pi/4$ (maximal mixing), in the  mass-mixing plane
$(\delta m^2,\sin^2\phi)$. All data are included in the fit (with standard 
{\em hep\/} flux.   It can be seen that both the LMA and LOW solutions are
consistent with maximal mixing (at 99\% C.L.)  for $\sin^2\phi\equiv
U^2_{e3}=0$. Moreever, the consistency of the LOW solution with maximal mixing
improves significantly for $U^2_{e3}\simeq 0.1$, while the opposite happens for
the LMA solution. This gives the possibility to obtain nearly bimaximal mixing
($\omega=\psi=\pi/4$ with $\phi$ small) within the LOW solution to the solar
neutrino problem---an interesting possibility for models predicting large
mixing angles.

\section{Summary}
\label{s5}

We have performed a thorough analysis of the  $2\nu$ and $3\nu$ MSW solutions
of the solar neutrino problem, discussing the information coming from  the
total rates, as well as from the SK energy spectrum and day-night asymmetry.
The global fit to the data puts an upper bound to the $\theta_{13}$ mixing
angle, consistent with atmospheric and reactor oscillation searches. A fraction
of the MSW solutions extends at and beyond maximal $(\nu_1,\nu_2)$ mixing
$(\theta_{12}\geq \pi/4)$, especially for the so-called LOW solution.
Therefore, a novel realization of (nearly) bimaximal mixing models appears
possible.

\acknowledgments

G.L.F.\ and E.L.\ thank the organizers of the {\em TAUP~'99\/} Workshop on
Topics in Astroparticle and Underground Physics (Paris, College de France),
where preliminary results of this work were presented. E.L.\ thanks  E.\ Kh.\
Akhmedov and P.\ I.\ Krastev for useful correspondence. This work is
co-financed by the Italian Ministero dell'Universit\`a e della Ricerca
Scientifica e Tecnologica (MURST) within the ``Astroparticle Physics'' project.

\appendix
\section*{Statistical analysis}
\label{app}

In this appendix we present an updated version of the $\chi^2$ statistical
analysis of neutrino event rates  discussed in Ref.~\cite{Fo95}, which has 
been widely used in many other analyses of the solar neutrino problem,
including the one performed by the Super-Kamiokande Collaboration in
\cite{SKDN}.  In particular, we take into account the most recent estimates for
the relevant BP98 standard solar model ingredients and for their uncertainties
\cite{BP98}.  Unless otherwise noted, the notation is the same as in
Ref.~\cite{Fo95}, to which the reader is referred for further details.

Table~\ref{tablefluxes} reports the solar neutrino fluxes $\phi_i$, as
calculated in the BP98 Standard Solar Model \cite{BP98}. The  corresponding
energy-averaged interaction cross sections $C_{ij}$ for the $j$-th detector are
given in Tab.~\ref{tablecross}, taking into account updated energy spectra
\cite{BP98,Al96} and cross sections \cite{Al96,GaCs,Ka95}. The associated
$1\sigma$ relative uncertainties $\Delta \ln C_{ij}$ are reported in
Tab.~\ref{tablecrosserr}. Such tables allow the calculation of the theoretical
uncertainties related to the detection process.

The uncertainties of the SSM $\nu$ fluxes  are mainly related on eleven basic
ingredients $X_k$: five astrophysical $S$-factors, the  Sun luminosity, the
metallicity $Z/X$, the Sun age, the opacity, the element diffusion, and the
$^7$Be electron capture rate $C_{\rm Be}$. The last two sources of
uncertainties represent new entries with respect to the list given in
\cite{Fo95}. The ``diffusion error'' is estimated by taking differences between
quantities calculated with and without diffusion \cite{BP98}. The uncertainty
in the  $^7$Be($e^-,\nu_e$)$^7$Li capture rate \cite{Gr97} affects (inversely)
the $^8$B flux by changing the $^7$Be density for the (much slower) competing
reaction  $^7$Be($p,\gamma$)$^8$B. The fractional $1\sigma$ uncertainties of
the eleven  input ingredients ($\Delta \ln X_k$) are reported in 
Tab.~\ref{tableinput}.

Table~\ref{tablealpha} reports the  matrix $\alpha_{ik}$ of logarithmic
derivatives  $\partial \ln \phi_i/\partial \ln X_k$, which parameterize the SSM
``response'' to small variations in the input ingredients, and which are
crucial to estimate correctly the correlations among the neutrino flux or event
rate uncertainties \cite{Fo95}. As remarked in \cite{Fo95} for the opacity
uncertainty, we have split the known flux errors  $\Delta \ln \phi_i=
\alpha_{ik}\, \Delta \ln X_k$ due to opacity, diffusion, and $^7$Be electron
capture, into two arbitrary factors, $\alpha_{ik}$ and $\Delta \ln X_k$ ($k=9$,
10, 11), so as to treat homogeneously all sources of uncertainties in the
$\chi^2$ statistics.

\begin{figure}
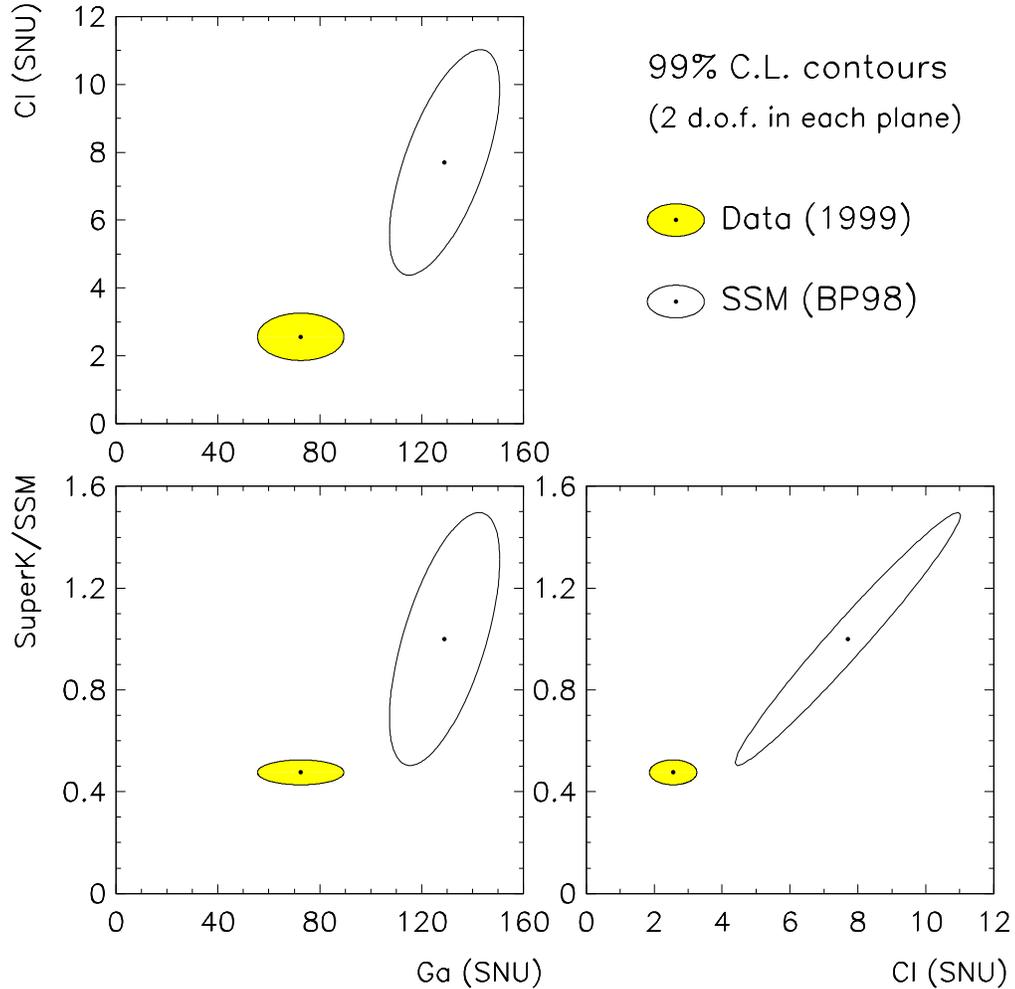

\caption{
The solar neutrino deficit, shown as a discrepancy between data and
expectations in the gallium (Ga), chlorine (Cl), and Super-Kamiokande total
event rates. In each plane, the error ellipses represent 99\% C.L.\ contours
for two degrees of freedom (i.e., $\Delta\chi^2=9.21$). The  projection of an
ellipse onto one of the axis gives approximately the $\pm3\sigma$ range for the
corresponding rate. Data and expectations refer to
Table~\protect\ref{tablerates}. The correlation of SSM errors is calculated as
in the Appendix.
}
\label{f1}
\end{figure}

\begin{figure}
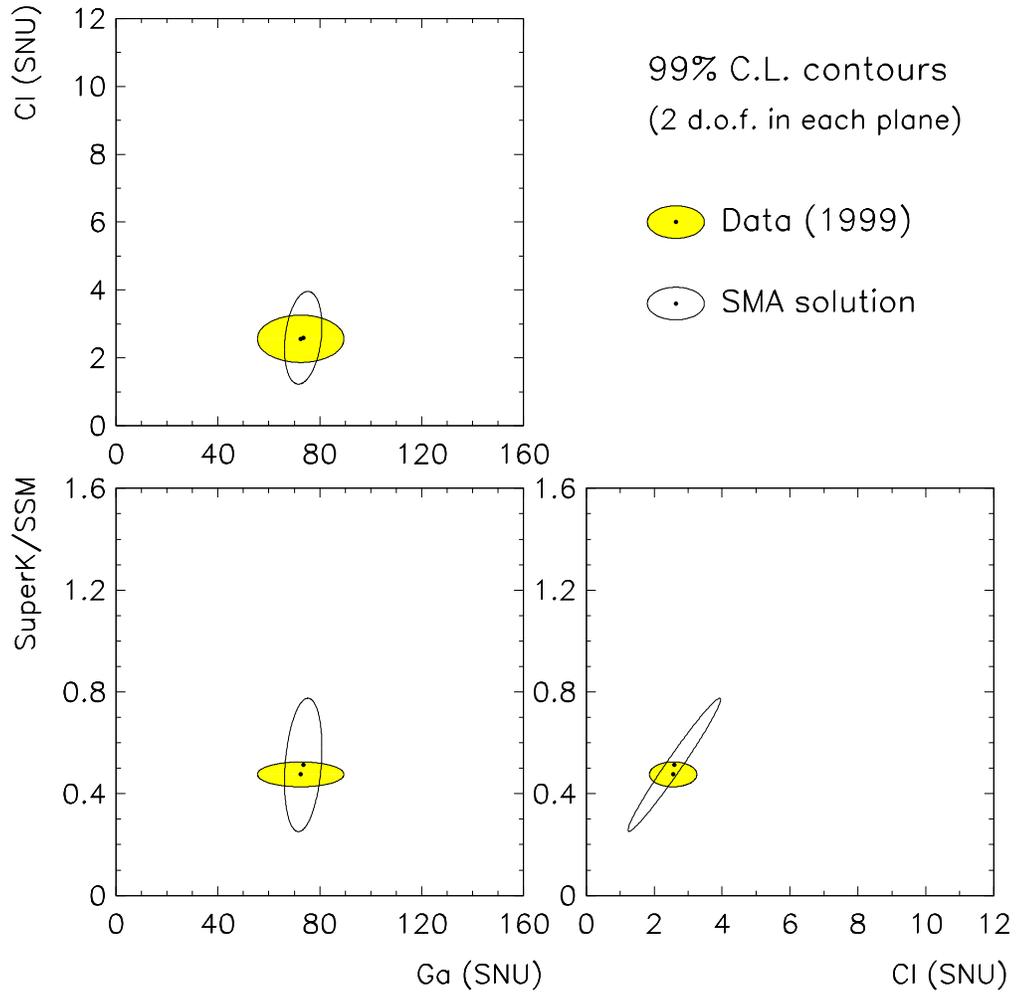

\caption{
The SMA solution at best fit (total rates only, first row of
Tab.~\protect\ref{chisquare}), compared with the experimental data, in the same
planes as in Fig.~\protect\ref{f1}. Note the excellent agreement between theory
and observations.
}
\label{f3}
\end{figure}

\begin{figure}
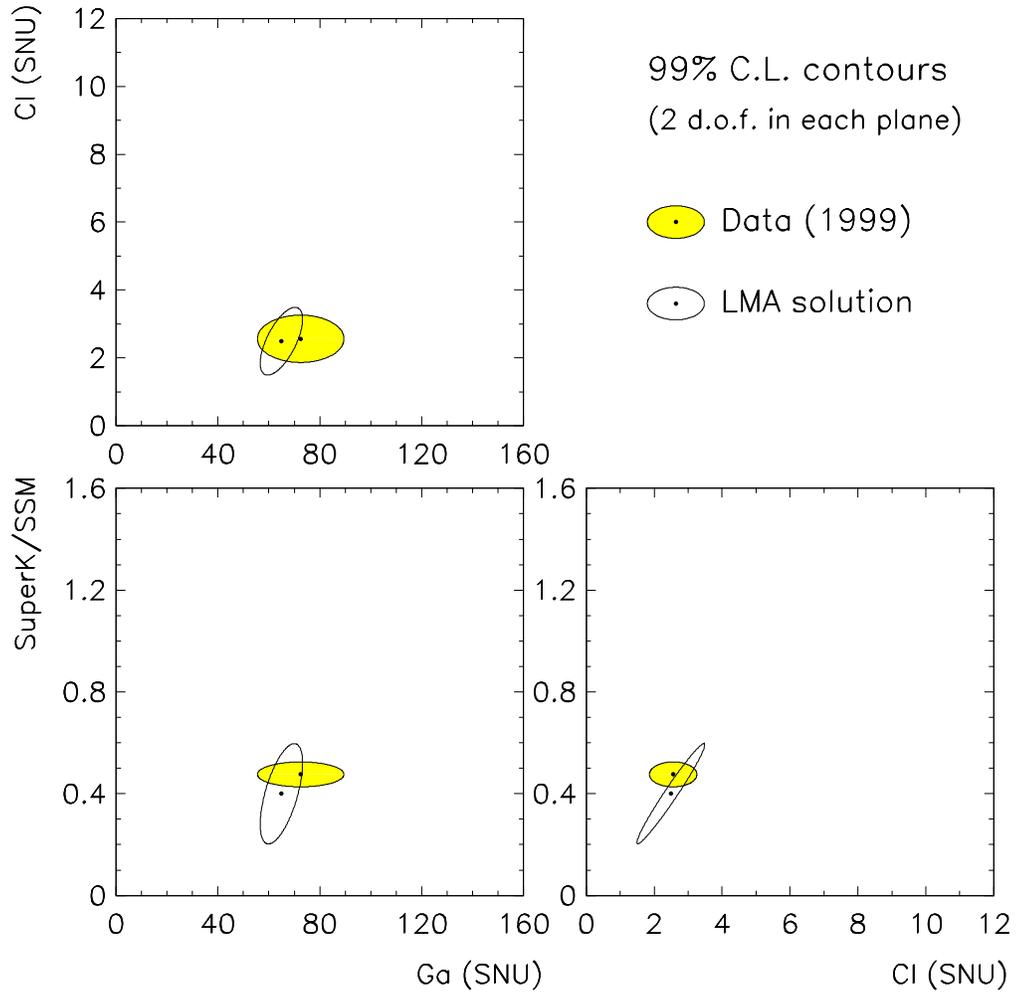

\caption{
As in Fig.~\protect\ref{f3}, but for the LMA solution at best fit  (total rates
only, second row of Tab.~\protect\ref{chisquare}).
}
\label{f4}
\end{figure}

\begin{figure}
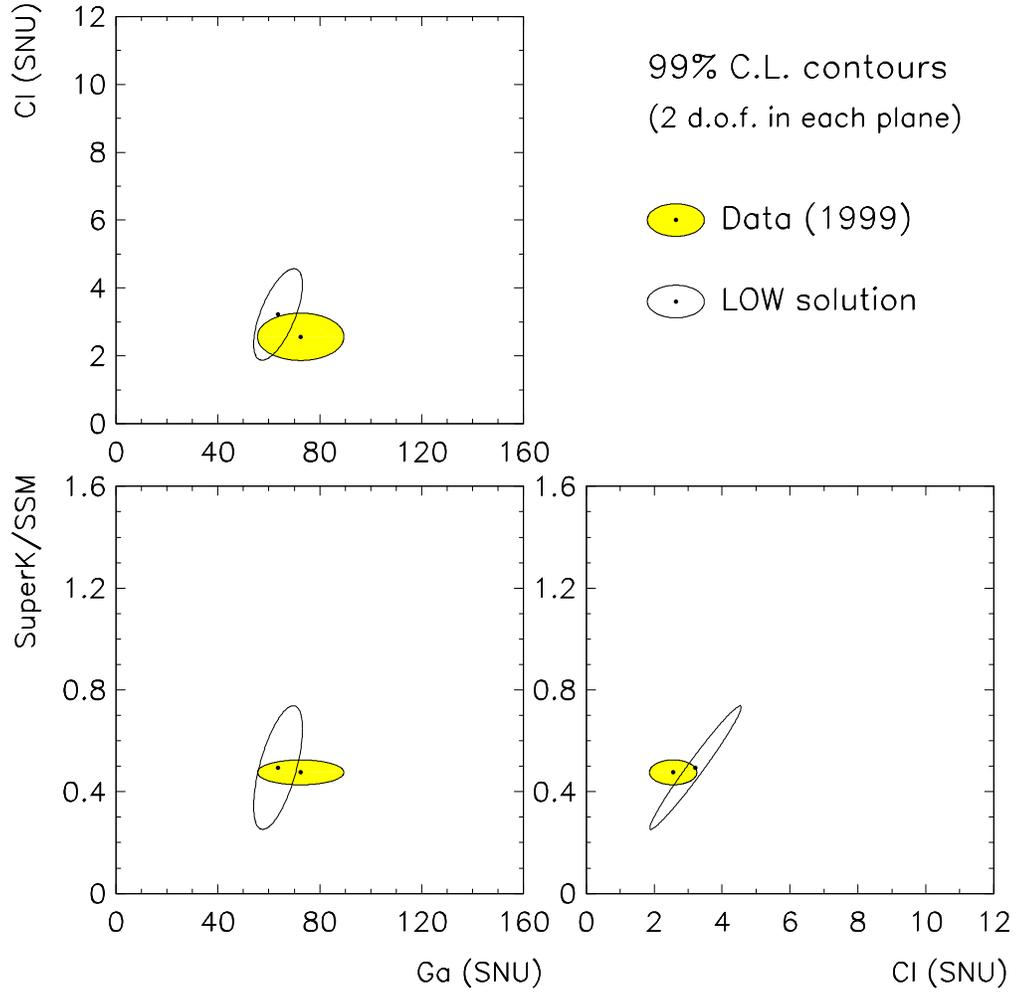

\caption{
As in Fig.~\protect\ref{f3}, but for the LOW solution at best fit  (total rates
only, third row of Tab.~\protect\ref{chisquare}).
}
\label{f5}
\end{figure}

Given the previous ingredients, the standard event rates $R_j$ are  calculated
as:
\begin{equation}
R_j = \sum_i R_{ij} = \sum_i \phi_i \, C_{ij}\ ,
\label{a1}
\end{equation}
and the associated theoretical error matrix ${\sigma^2_{j_1 j_2}}$ reads:
\begin{equation}
\sigma^2_{j_1 j_2} = \delta_{j_1 j_2}\sum_i R^2_{i j_1}\,(\Delta C_{i j_1})^2
+ \sum_{i_1,i_2} R_{i_1 j_1}\, R_{i_2 j_2} 
\sum_k \alpha_{i_1 k}\,\alpha_{i_2 k}\,(\Delta \ln X_k)^2\ .
\label{a2}
\end{equation}
The off-diagonal entries of ${\sigma^2_{j_1 j_2}}$ induce strong correlations
among the theoretical uncertainties for different experiments, as evident in
Fig.~\ref{f1}. In the presence of oscillations  (Figs.~\ref{f3}--\ref{f5}), 
the standard values of the partial rates $R_{ij}$ are simply replaced  by the
corresponding oscillated values.

We remind that the entries of the $\alpha_{ik}$ matrix are not independent,
being constrained by the luminosity sum rule \cite{Fo95}. At equilibrium, the
standard neutrino fluxes must obey the energy conservation relation
\cite{NuAs}:
\begin{equation}
\sum_i (Q/2 - E_i) \, \phi_i = K\ ,
\label{a3}
\end{equation}
where $E_i$ is the average neutrino energy for the $i$-th neutrino flux,
$K=0.853 \times 10^{12}$ MeV~cm$^{-2}$~s$^{-1}$ is the solar luminosity, and
$Q=26.73$ MeV is the overall $Q$-value for the fusion reaction  $4p\to
\alpha+2\nu_e+2e^+$ (independent from the intermediate reaction chain).

Non-equilibrium corrections, which are especially relevant for CNO fluxes,
modify this relation into \cite{Ul96,BaKr}
\begin{equation}
\sum_i \xi_i \, \phi_i = K\ ,
\label{a4}
\end{equation}
where the $\xi_i$ values are given in Tab.~\ref{tableluminosity}. By taking
partial derivatives of the above relation with respect to the input ingredients
$X_k$, one gets the sum rule \cite{Fo95}:
\begin{equation}
\frac{\sum_i \xi_i \, \phi_i \, \alpha_{ik}}{\sum_i \xi \,\phi_i} = 
\delta_{k6}\ ,
\label{a5}
\end{equation}
which is satisfied with good approximation ($\sim2\%$ level).

Similarly, if the flux variations are approximately parametrized  as power laws
of the central central temperature $T$ of the Sun,
\begin{equation}
\phi_i(T)= \phi_i(T_0)\,\left(\frac{T}{T_0} \right)^{\beta_i}\ ,
\label{a6}
\end{equation}
the temperature exponents $\beta_i$ become constrained by the sum rule
\cite{Fo95,Ul96}
\begin{equation}
\frac{\sum_i \xi_i\, \beta_i \, \phi_i}{\sum_i \xi_i \phi_i}=0
\label{a7}
\end{equation}
which, using the exponents in Ref.~\cite{Ul96}, is satisfied only at the $\sim
10\%$ level. However,  such exponents can be easily adjusted within their
uncertainties $(\beta_i\to \tilde \beta_i)$ so as to  satisfy exactly the above
sum rule. Sets of temperature exponents are given in  Tab.~\ref{tablebeta}. The
set of $\tilde \beta_i$'s is particularly relevant for analyses of the solar
neutrino problem with ``unconstrained''  central temperature (not performed in
this work).

Since the above tables and equations refer only to the analysis of the {\em
total\/}  neutrino event rates, a final remark is in order about the fit to the
SK  energy-angle spectral information. For MSW analyses, the maximum amount of
information about the SK solar neutrino signal is contained in the
double-differential distribution of events in nadir angle  $(\eta)$ and
electron recoil energy $(E_e)$. However, the $(\eta,\,E_e)$ distribution is not
yet available outside the SK collaboration, which has released only the two
{\em projected\/} distributions, namely, the (angle-averaged) energy spectrum
and the (energy-averaged) nadir distribution. The experimental errors of such
projections (including statistical fluctuations) are {\em a priori\/} strongly
correlated, but the published information is not sufficient to recover such
correlations. For such reasons, while we include the complete information in
$E_e$ above 5.5 MeV (given as an 18-bin spectrum in \cite{To99,Su99,Na99}), we
prefer  to use only a minimal (but relevant) information in $\eta$, namely, 
the night-day asymmetry $(N-D)/(N+D)$.

The $\chi^2$ function for the energy spectrum is then constructed in the same
way as in Ref.~\cite{800d}, namely, by separating uncorrelated and correlated
error components in the $18\times 18$ error matrix, with off-diagonal
correlations equal to $+1$. A numerical table of errors is explicitly reported
in \cite{800d} for the 825 day SK spectrum. Double counting of the total rate
information in SK is avoided by minimizing the $\chi^2$  with respect to the
overall spectrum normalization, taken as a free parameter \cite{800d}. In the
global fit, we then sum up the $\chi^2$ of the spectrum,  the $\chi^2$ of the
total rates, and the (trivial) $\chi^2$ of the day-night asymmetry measurement.
A similar construction of the global $\chi^2$ function  is also adopted in
\cite{SNOL}.


\begin{table}
\caption{Solar neutrino event rates observed in the five experiments Homestake,
	Kamiokande, SAGE, GALLEX, and Super-Kamiokande (825 days,  $E_e >6.5$
	MeV),  as compared with the theoretical expectations from the BP'98
	standard solar model \protect\cite{BP98}. The quoted uncertainties are
	at 1$\sigma$. In our analysis, the GALLEX and SAGE experimental rates
	are combined in quadrature in a single (Gallium) rate of $72.5\pm 5.6$
	SNU.}
\smallskip
\begin{tabular}{lcccc}
Experiment 	& Experimental rate 			& Theoretical rate 
& Units 		& Ref. 						\\
\tableline
Homestake 	& $2.56\pm0.16\pm0.16$ 			&$7.7\;^{+1.2}_{-1.0}$ 
& SNU 				& \protect\cite{Cl98} 			\\
Kamiokande 	& $2.80\pm0.19\pm0.33$ 			&$5.15\;^{+1.0}_{-0.7}$
& $10^6$ cm$^{-2}$~s$^{-1}$ 	& \protect\cite{Fu96} 			\\
SAGE 		& $67.2\;^{+7.2}_{-7.0}\;^{+3.5}_{-3.0}$&$129\;^{+8}_{-6}$
& SNU 				& \protect\cite{Ab99} 			\\
GALLEX 		& $77.5\pm6.2\;^{+4.3}_{-4.7}$ 		&$129\;^{+8}_{-6}$ 
& SNU 				& \protect\cite{Ha99} 			\\
Super-Kamiokande& $2.45\pm0.04\pm0.07$ 			&$5.15\;^{+1.0}_{-0.7}$
& $10^6$ cm$^{-2}$~s$^{-1}$ 	& \protect\cite{Su99} 
\end{tabular}
\label{tablerates}
\end{table}

\begin{table}
\caption{$2\nu$ MSW analysis: Absolute $\chi^2$ values in selected points of
	the parameter space $(\delta m^2,\sin^2 2\omega/\cos 2\omega)$,
	corresponding to  local $\chi^2$ minima (SMA, LMA, LOW solutions)  and
	to the no-oscillation case. Upper three rows: fit to total rates only
	(Cl+Ga+K+SK data). Middle three rows: fit to total rates (4 data),
	day-night asymmetry (1 datum), and SK energy spectrum (18 data with one
	adjustable normalization parameter). Lower three rows: as for the
	middle rows, but with {\em hep\/} flux increased by a factor 20. The
	number of degrees of freedom $N_{\rm DF}$ is equal to the number of
	data (either 4 or $4+1+18-1$) minus 2 (the oscillation parameters). The
	last column corresponds to the probability of having a worse $\chi^2$
	fit {\em a priori}.}
\smallskip
\begin{tabular}{l|ccccc}
Observables
&Solution&$\delta m^2$ (eV$^2$)&$\sin^22\omega/\cos2\omega$&$\chi^2$	   
& $P(\chi^2,N_{\rm DF})$\\
\tableline
 			& SMA &$5.89\times10^{-6}$&$5.62\times10^{-3}$&1.08
&58.3~\%\\
Rates only		& LMA &$1.91\times10^{-5}$&1.62		      &4.59
&10.1~\%\\
($N_{\rm DF}=2$)	& LOW &$1.07\times10^{-7}$&3.48	    	      &8.24
&1.6~\%\\
		    	& no osc. &0 		  &0		      &60.1
&$9\times10^{-14}$\\
\tableline
Rates,			& SMA &$5.37\times10^{-6}$&$6.50\times10^{-3}$&27.3
&12.7~\%\\
$N-D/N+D$,		& LMA &$2.55\times10^{-5}$&1.62		      &27.8
&11.4~\%\\
spectrum 		& LOW &$1.07\times10^{-7}$&2.87		      &30.8
&5.8~\%\\
($N_{\rm DF}=20$)& no osc. &0 		  &0		      &85.8
&$4\times 10^{-10}$\\
\tableline
Rates,			& SMA &$5.20\times10^{-6}$&$6.20\times10^{-3}$&25.6
&17.9~\%\\
$N-D/N+D$,		& LMA &$2.94\times10^{-5}$&1.96		      &24.6
&21.7~\%\\
spectrum [{\em hep}$\times20$]&LOW&$1.24\times10^{-7}$&2.87	      &27.8
&11.4~\%\\
($N_{\rm DF}=20$)	& no osc. &0 		  &0		      &71.9
&$9\times 10^{-8}$\\
\end{tabular}
\label{chisquare}
\end{table}

\begin{table}
\caption{$3\nu$ MSW analysis: Absolute minima of the function $\chi^2(\delta
	m^2,\tan^2\omega,\tan^2\phi)$ and their associated  probability $P$.
	The number of degrees of freedom is $N_{\rm DF}=1$ for the fit to total
	rates only and $N_{\rm DF}=19$ for the global fit.}
\smallskip
\begin{tabular}{l|ccccc}
Observables & $\delta m^2$ (eV$^2$) & $\tan^2\omega$ & $\tan^2\phi$ & 
$\chi^2_{\min}$ & $P(\chi^2,N_{\rm DF})$ \\
\tableline
Rates only			& $9.8\times10^{-6}$ & $8.1\times10^{-4}$ &
0.1 & 0.70 & 40.3~\% \\
All data			& $9.8\times10^{-6}$ & $7.1\times10^{-4}$ &
0.1 & 27.0 & 10.5~\% \\
All data [{\em hep}$\times 20$]	& $2.8\times10^{-5}$ & $0.37$             &
0   & 24.5 & 17.8~\%
\end{tabular}
\label{chiminima}
\end{table}

\begin{table}
\caption{Neutrino fluxes $\phi_i$ (cm$^{-2}$~s$^{-1}$) from the BP98 SSM
\protect\cite{BP98}.}
\smallskip
\begin{tabular}{cccccccc}
pp 	& pep 	& hep 	& Be 	& B 	& N 	& O 	& F 		\\
$\times10^{10}$ & $\times10^8$ 	& $\times10^3$ 	& $\times10^9$ 		& 
$\times10^6$    & $\times10^8$ 	& $\times10^8$ 	& $\times10^6$ 		\\
\tableline
5.94 	& 1.39 	& 2.10 	& 4.80 	& 5.15 	& 6.05 	& 5.32 	& 6.33 
\end{tabular}
\label{tablefluxes}
\end{table}

\begin{table}
\caption{Energy-averaged cross sections $C_{ij}$. Units are such that the
	expected partial rates $R_{ij}=C_{ij}\phi_i$ are directly given in SNU
	for the Ga and Cl experiments, and as a ratio to the total BP98
	predicted rate for Super-Kamiokande ($E_e>6.5$ MeV) and Kamiokande.}
\smallskip
\begin{tabular}{lcccccccc}
	&pp 	& pep 	& hep 	& Be 	& B 	& N 	& O 	& F 	  \\
	&$\times10^{-9}$ &$\times10^{-8}$&$\times10^{-6}$& $\times10^{-9}$& 
	$\times10^{-6}$ &$\times10^{-9}$&$\times10^{-8}$& $\times10^{-8}$\\
\tableline
Ga	& 1.17	& 2.04	& 7.14	& 7.17	& 2.40	& 6.04	& 1.14	& 1.14	\\
Cl	& 0	& 0.16	& 4.26	& 0.24	& 1.14	& 0.17	& 0.07	& 0.07	\\
K	& 0	& 0	& 0.595	& 0	& 0.194	& 0	& 0	& 0	\\ 
SK	& 0	& 0	& 0.514	& 0	& 0.194 & 0	& 0	& 0	
\end{tabular}
\label{tablecross}
\end{table}

\begin{table}
\caption{$1\sigma$ relative errors  $\Delta \ln C_{ij}$ of the energy averaged
	cross sections. The uncertainties of the (radiatively corrected)
	$\nu$-$e$ scattering cross section in K and SK are negligible.}
\smallskip
\begin{tabular}{lcccccccc}
	&pp 	& pep 	& hep 	& Be 	& B 	& N 	& O 	& F 	  \\
\tableline
Ga	& 0.023	& 0.170	& 0.320	& 0.070	& 0.320	& 0.060	& 0.120	& 0.120	  \\
Cl	& 0	& 0.020	& 0.037	& 0.020	& 0.032	& 0.020	& 0.020	& 0.020	  \\
K, SK	& 0	& 0	& 0	& 0	& 0	& 0	& 0	& 0	
\end{tabular}
\label{tablecrosserr}
\end{table}

\begin{table}
\caption{$1\sigma$ relative errors  $\Delta \ln X_k$ of the relevant SSM input
	parameters.}
\smallskip
\begin{tabular}{ccccccccccc}
$S_{11}$&$S_{33}$&$S_{34}$&$S_{1,14}$&$S_{17}$&Lum&$Z/X$&Age&Opa&Diff&
$C_{\rm Be}$\\
\tableline
0.017 &0.060 &0.094 &0.143 &0.106 &0.004 &0.033 &0.004 &0.02 &0.02 & 0.02
\end{tabular}
\label{tableinput}
\end{table}

\begin{table}
\caption{Logarithmic derivatives $\alpha_{ik}=\partial \ln \phi_i/\partial  \ln
	X_k$, parameterizing the response of neutrino fluxes $\phi_i$ to
	variations in the SSM input parameters $X_k$.}
\smallskip
\begin{tabular}{lcccccccc}
	     &pp     & pep   & hep   & Be    & B     & N     & O     & F     \\
\tableline
$S_{11}$     &$+0.14$&$-0.17$&$-0.08$&$-0.97$&$-2.59$&$-2.53$&$-2.93$&$-2.94$\\
$S_{33}$     &$+0.03$&$+0.05$&$-0.45$&$-0.43$&$-0.40$&$+0.02$&$+0.02$&$+0.02$\\
$S_{34}$     &$-0.06$&$-0.09$&$-0.08$&$+0.86$&$+0.81$&$-0.05$&$-0.05$&$-0.05$\\
$S_{1,14}$   &$-0.02$&$-0.02$&$-0.01$&$+0.00$&$+0.01$&$+0.85$&$+1.00$&$+0.01$\\
$S_{17}$     &$+0.00$&$+0.00$&$+0.00$&$+0.00$&$+1.00$&$+0.00$&$+0.00$&$+0.00$\\
Lum	     &$+0.73$&$+0.87$&$+0.12$&$+3.40$&$+6.76$&$+5.16$&$+5.94$&$+6.25$\\
$Z/X$	     &$-0.08$&$-0.17$&$-0.22$&$+0.58$&$+1.27$&$+1.86$&$+2.03$&$+2.09$\\
Age	     &$-0.07$&$+0.00$&$-0.11$&$+0.69$&$+1.28$&$+1.01$&$+1.27$&$+1.29$\\
Opa	     &$+0.14$&$+0.24$&$+0.54$&$-1.38$&$-2.62$&$-1.67$&$-2.05$&$-2.13$\\
Diff	     &$+0.13$&$+0.22$&$+0.13$&$-0.90$&$-2.00$&$-2.56$&$-2.75$&$-2.75$\\
$C_{\rm Be}$ &$+0.00$&$+0.00$&$+0.00$&$+0.00$&$-1.00$&$+0.00$&$+0.00$&$+0.00$
\end{tabular}
\label{tablealpha}
\end{table}

\begin{table}
\caption{Neutrino energy parameters (in MeV) relevant for the luminosity sum
	rules [Eqs.~(\protect\ref{a3}--\protect\ref{a5})].}
\smallskip
\begin{tabular}{lcccccccc}
	     &pp     & pep   & hep   & Be    & B     & N     & O     & F     \\
\tableline
$E_i$        &$0.261$&$1.442$&$9.625$&$0.813$&$6.735$&$0.706$&$0.994$&$1.000$\\
$Q/2-E_i$    & 13.10 & 11.92 & 3.74  & 12.55 & 6.63  & 12.66 & 12.37 & 12.37 \\
$\xi_i$      & 13.10 & 11.92 & 10.17 & 12.55 & 6.63  & 3.457 & 21.57 & 2.36 
\end{tabular}
\label{tableluminosity}
\end{table}

\begin{table}
\caption{Temperature exponents $\beta_i$ and their $\pm 1\sigma$ errors (from
	\protect\cite{Ul96,NuAs}).  The exponents $\tilde\beta_i$ represent our
	best-fit adjustments, which satisfy exactly the luminosity sum rule
	[Eq.~(\protect\ref{a7})].}
\smallskip
\begin{tabular}{lcccccccc}
	     &pp     & pep   & hep   & Be    & B     & N     & O     & F\\
\tableline
$\beta_i$    &$-1.1\pm0.1$&$-2.4\pm0.9$&$4.5\pm1.5$&$10\pm2$&$24\pm5$&
$24.4\pm0.2$&$27.1\pm0.1$&$27.8\pm0.1$\\
$\tilde\beta_i$ &$-1.1039$ & $-2.407$ & 4.5 &8.79&23.996&24.4&27.099&27.8 \\ 
\end{tabular}
\label{tablebeta}
\end{table}



\begin{figure}
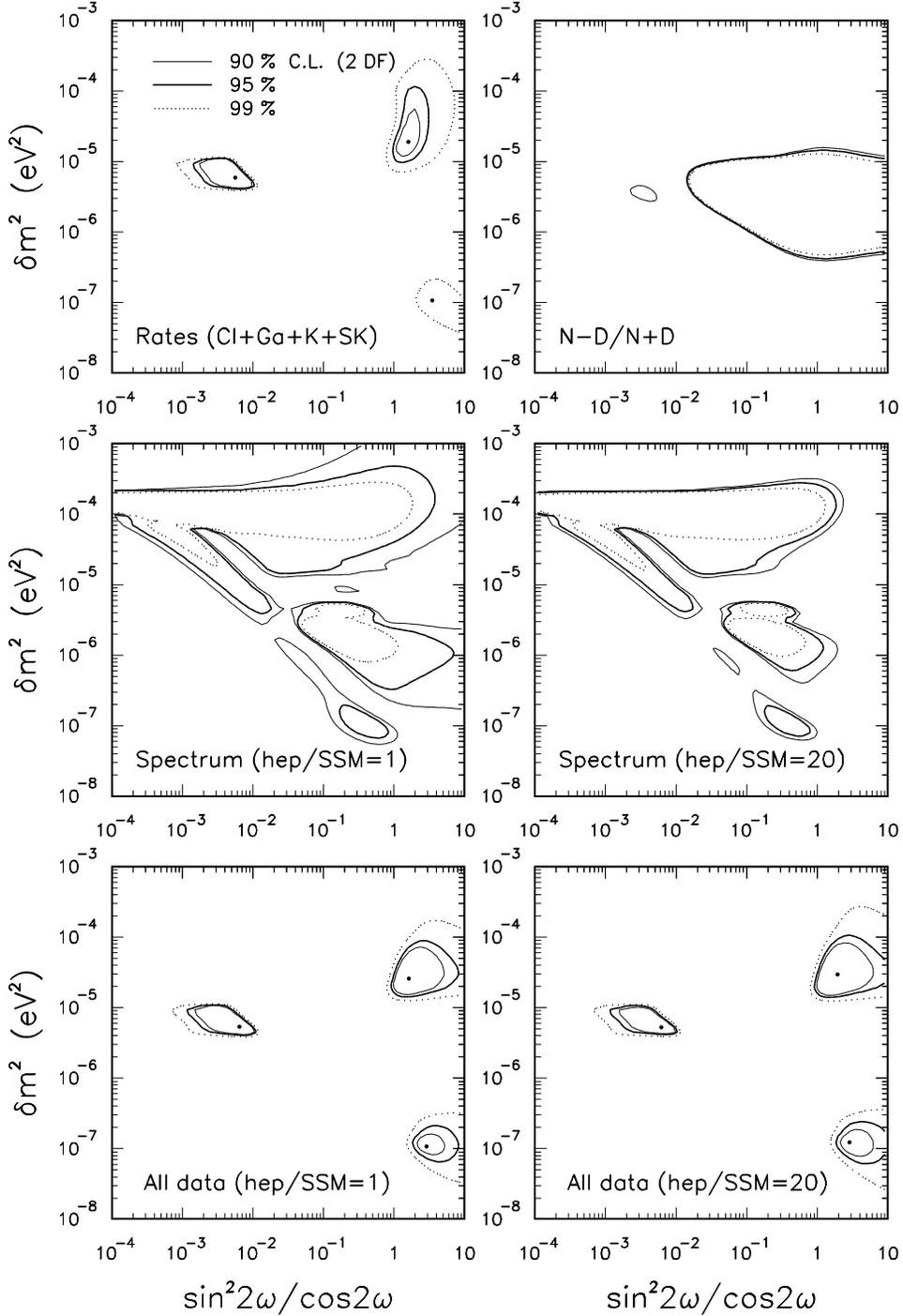

\caption{
Two-generation MSW solutions to the solar neutrino problem. The upper  four
panels correspond to the following separate fits to data subsets:  total rates
(Cl+Ga+K+SK); Super-Kamiokande night-day asymmetry $N-D/N+D$; Super-Kamiokande
electron energy spectrum with standard {\em hep\/} neutrino flux;
Super-Kamiokande spectrum with enhanced $(20\times)$ {\em hep\/} neutrino flux.
The two lower panels show the results of global fits to all data. The thin
solid, thick solid, and dashed curves correspond to $\chi^2-\chi^2_{\rm
min}=4.61$, 5.99, and 9.21. The positions of the local $\chi^2$ minima in fits
including the total rates are indicated by dots. See also
Tab.~\protect\ref{chisquare}.
}
\label{f2}
\end{figure}
\begin{figure}
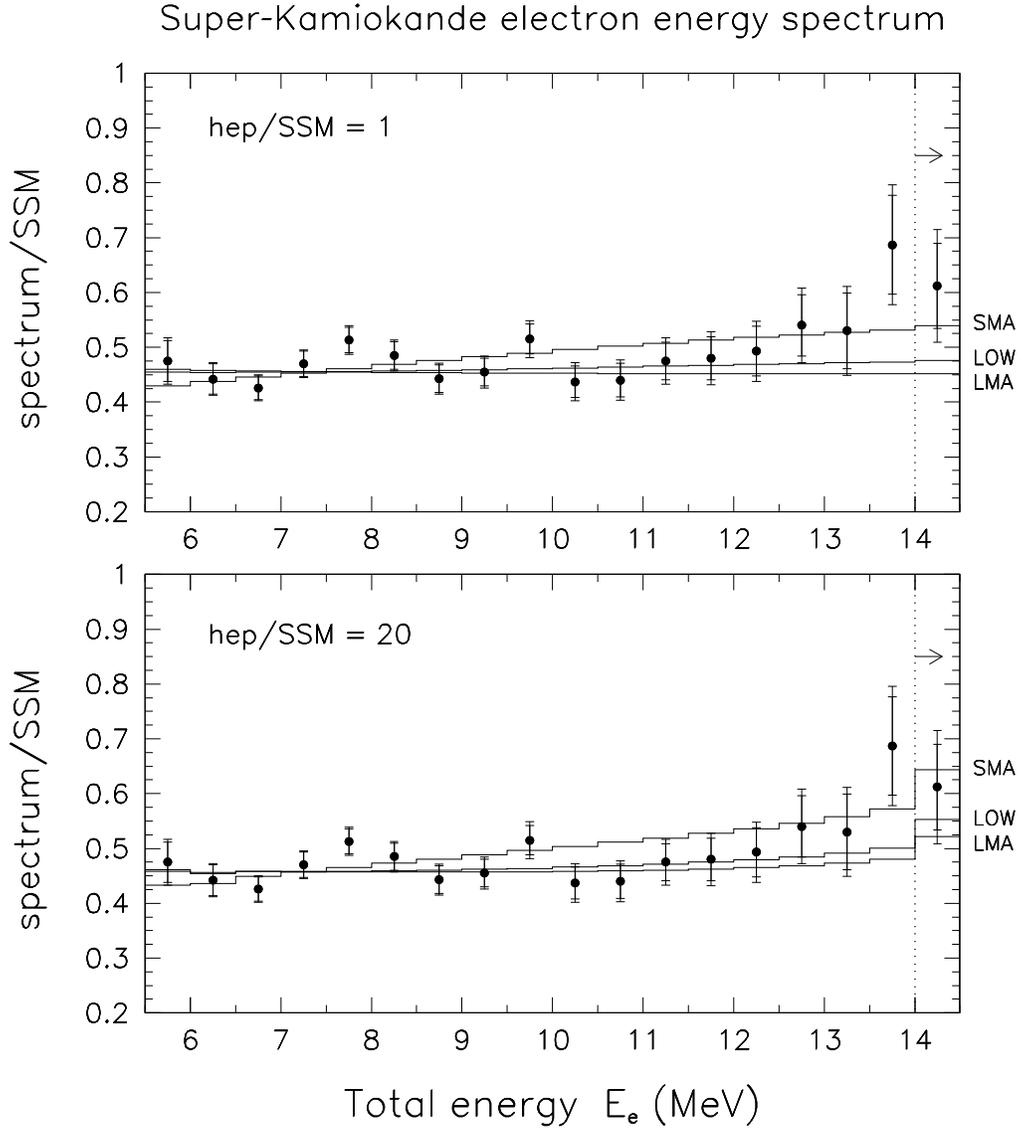

\caption{
Comparison of data and predictions for the Super-Kamiokande spectrum shape. The
theoretical spectra normalization is taken free in the fit. The upper (lower)
panel corresponds to the case of standard  ($20\times$) {\em hep\/} flux. The
SMA, LMA, and LOW spectra are calculated in the global best fit points reported
in  Tab.~\protect\ref{chisquare} (middle rows for the upper panel and lower
rows for the lower panel). The SK data are reported from 
\protect\cite{To99,Su99,Na99}, and the error bars refer to statistical and
total experimental errors.
}
\label{f6}
\end{figure}
\begin{figure}
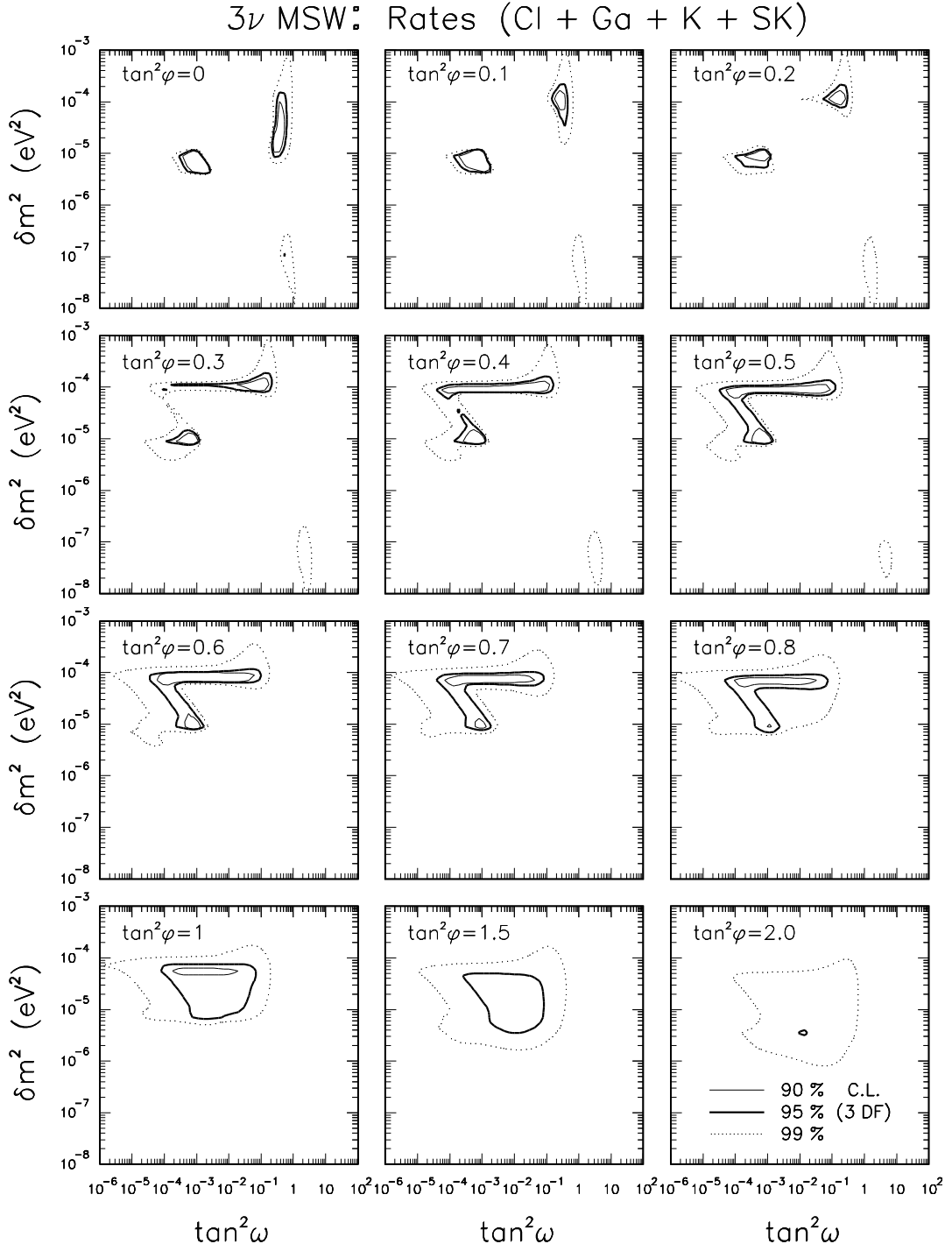

\caption{
Three-flavor MSW oscillations: global fit to Cl+Ga+K+SK rates in the $(\delta
m^2,\tan^2\omega,\tan^2\phi)$ parameter space. The favored regions in each
panel correspond to sections of the volume  allowed at 90\%, 95\%, and 99\%
C.L.\  ($\chi^2-\chi^2_{\min}=6.25$, 7.82, and 11.36) for representative values
of $\tan^2\phi$. 
}
\label{f7}
\end{figure}
\begin{figure}
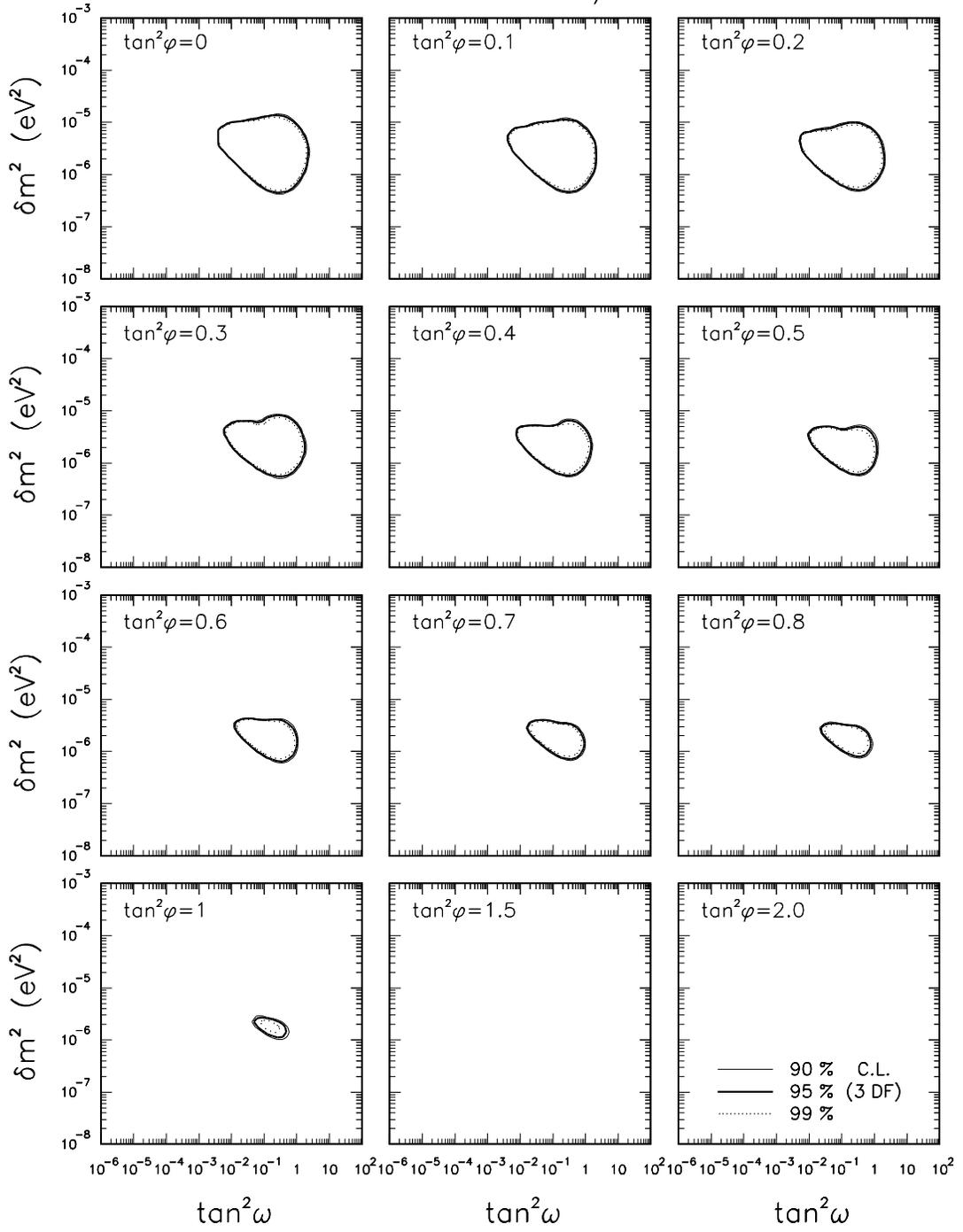

\caption{
As in Fig.~\protect\ref{f7}, but for the fit to the Super-Kamiokande night-day
asymmetry. The region inside the curves is excluded.
}
\label{f8}
\end{figure}
\begin{figure}
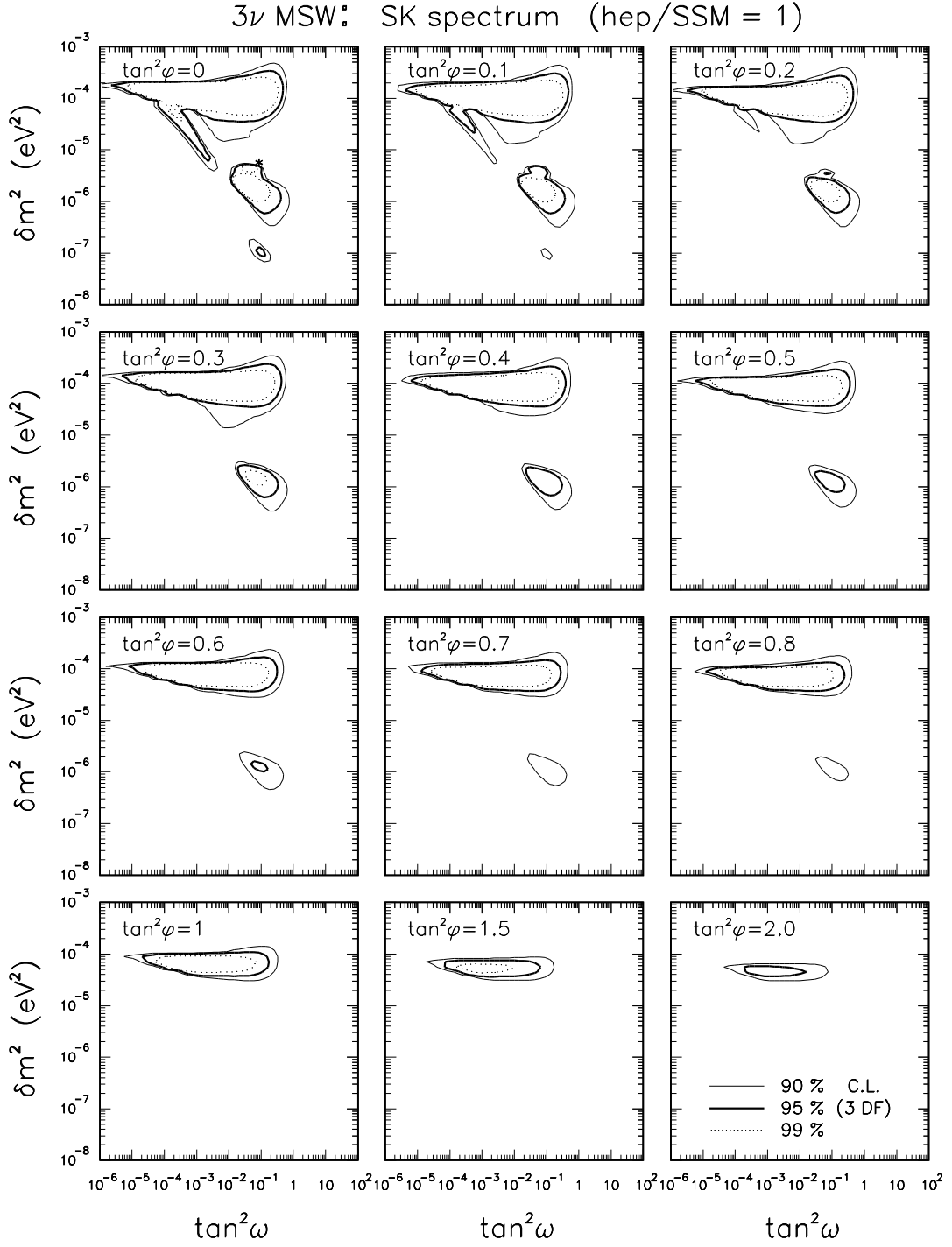

\caption{
As in Fig.~\protect\ref{f7}, but for the fit to the Super-Kamiokande energy
spectrum. The regions inside the curves are excluded.
}
\label{f9}
\end{figure}
\begin{figure}
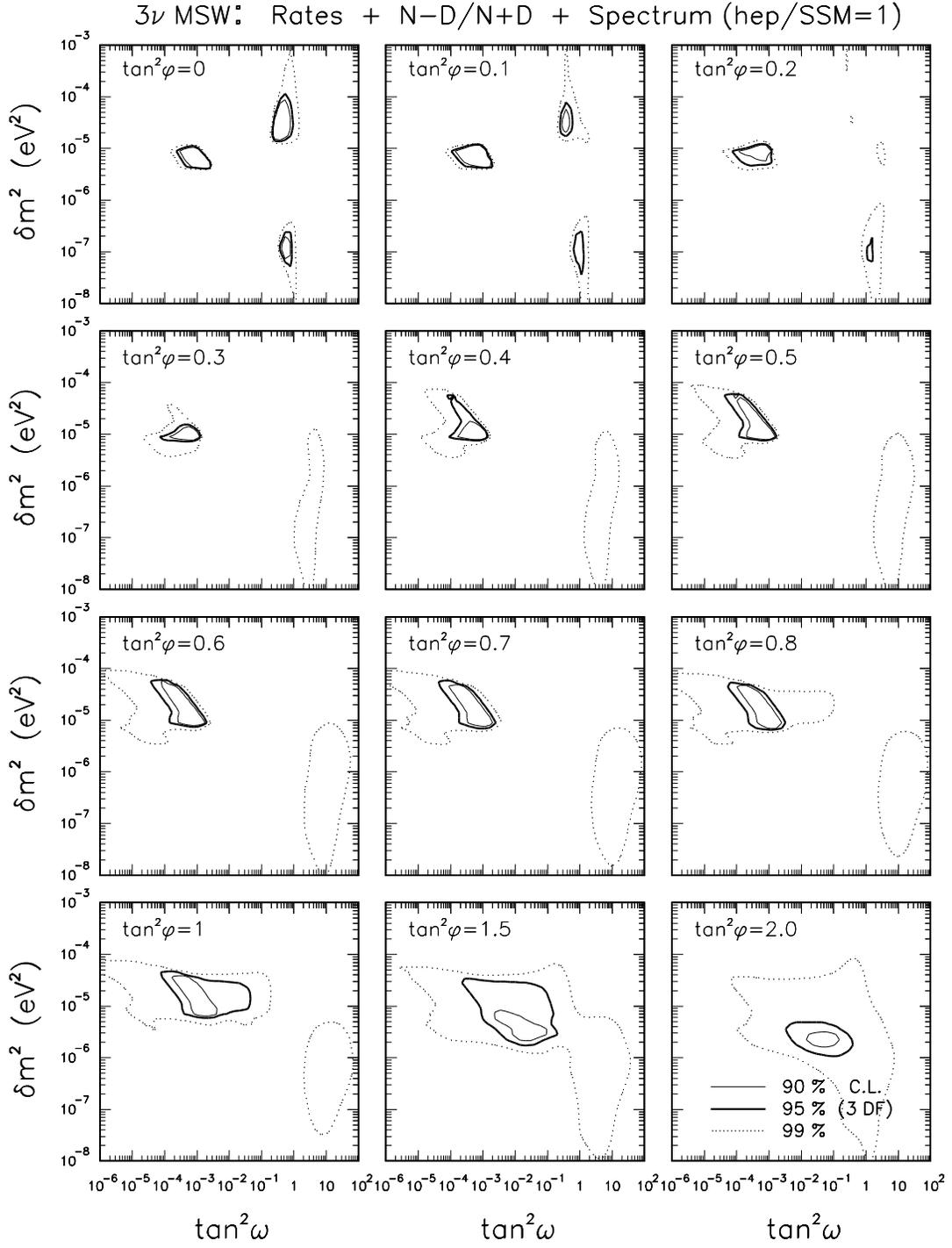

\caption{
Results of the global three-flavor MSW fit to all data. Notice that, in the
first two panels, the 99\% C.L.\  contours  are compatible with maximal mixing
($\tan^2\omega=1$) for both the LOW and the LMA solutions.
}
\label{f10}
\end{figure}
\begin{figure}
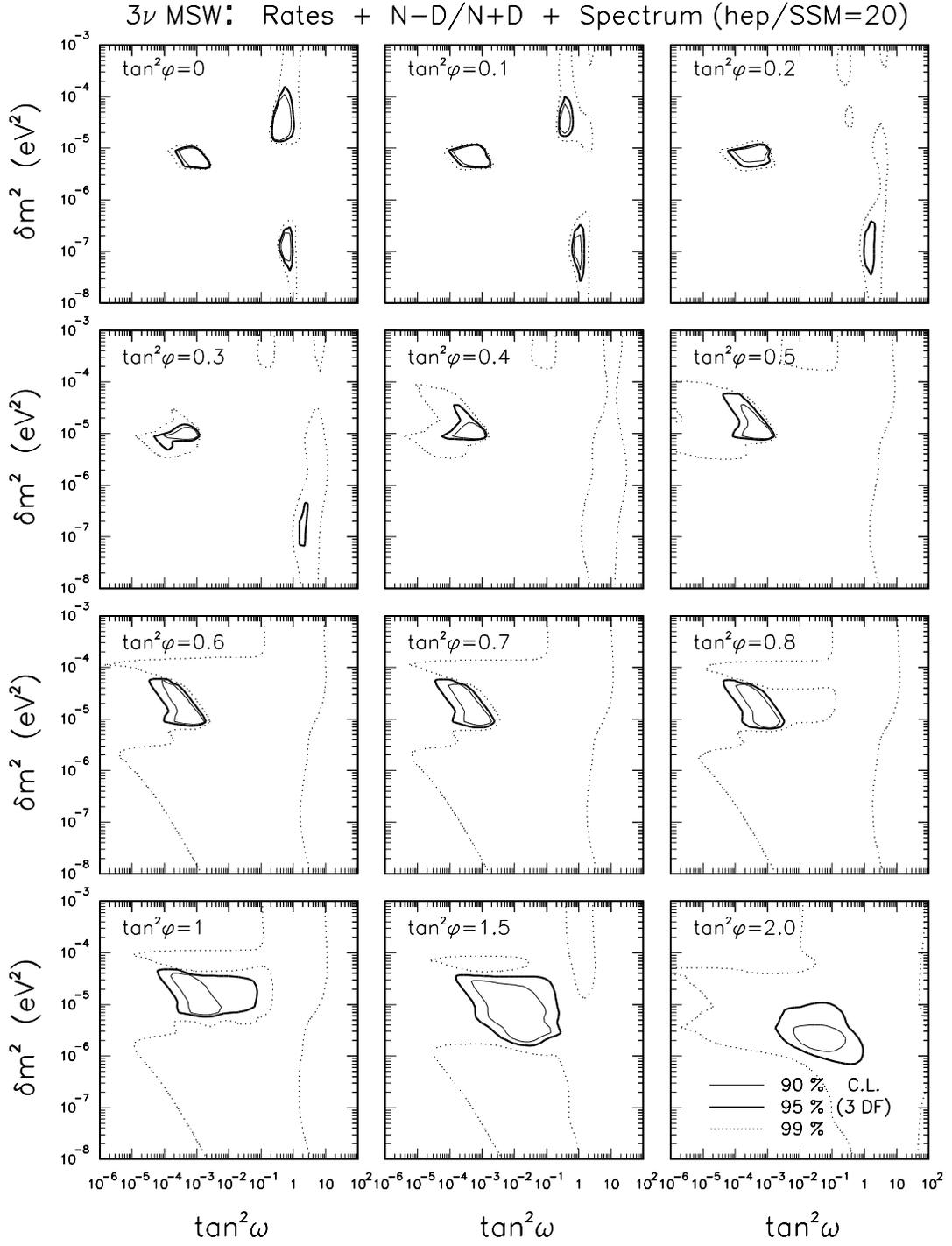

\caption{
As in Fig.~\protect\ref{f10}, but for the case of enhanced ($20\times$) {\em
hep\/} flux. The allowed regions are slightly enlarged with respect to
Fig.~\protect\ref{f10}.
}
\label{f11}
\end{figure}
\begin{figure}
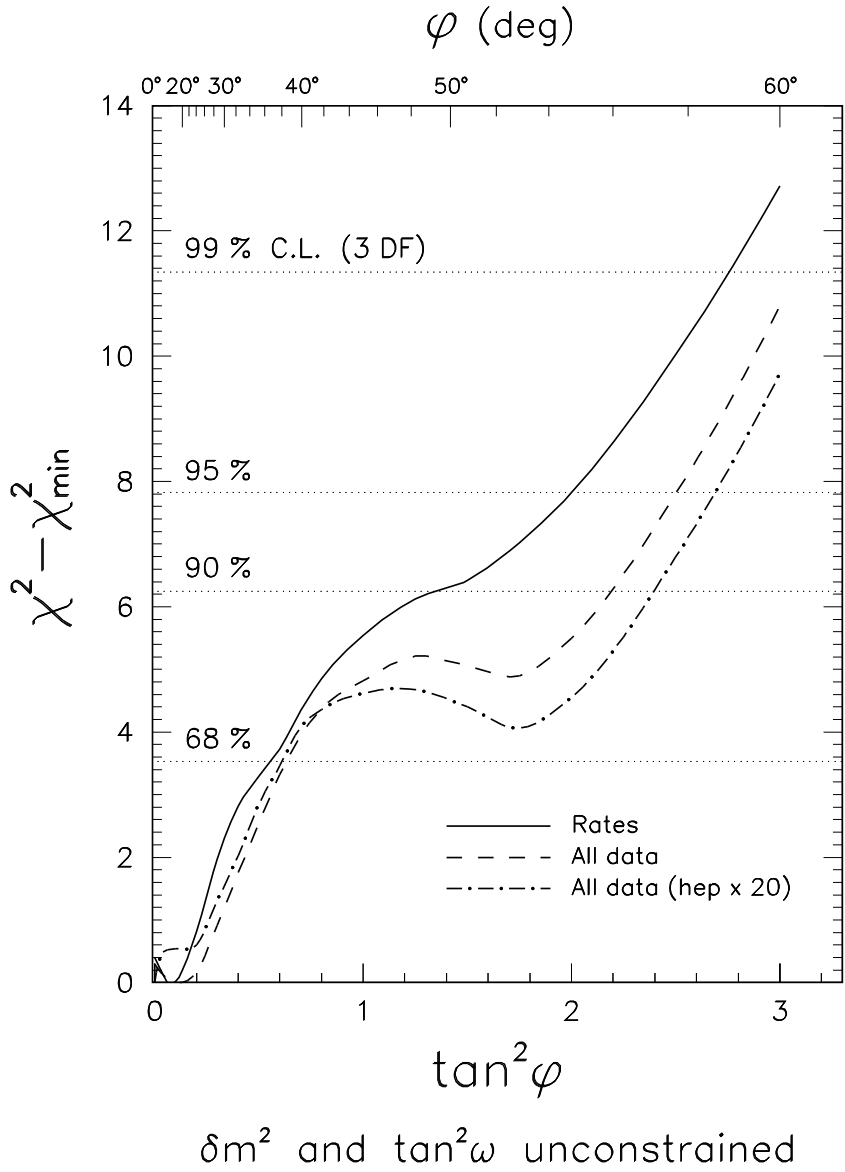

\caption{
Values of $\Delta\chi^2$ as a function of $\phi$, for unconstrained $\delta
m^2$ and $\tan^2\omega$. At 95\% C.L., the upper limit on $\phi$  is in the
range $55^\circ$--$59^\circ$, depending on the data used in the fit and on the
value of the {\em hep\/} flux.
}
\label{f12}
\end{figure}
\begin{figure}
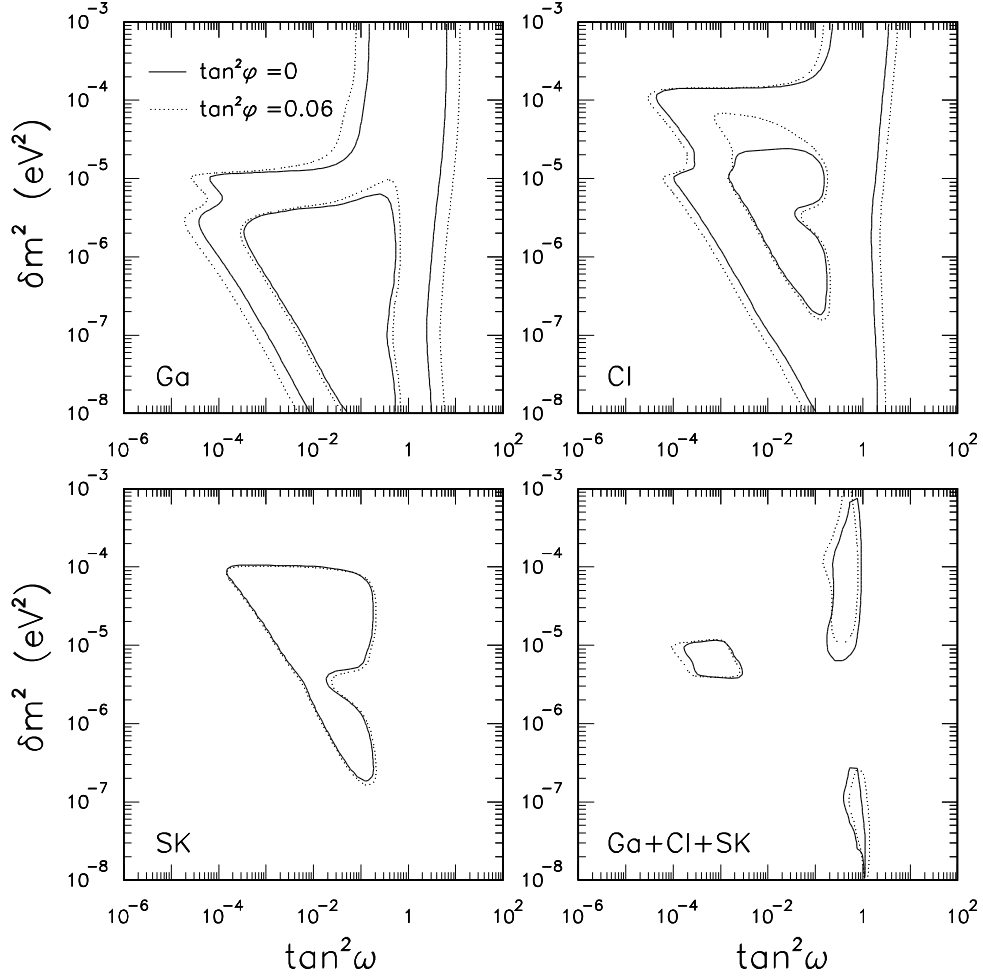

\caption{
Regions allowed at 99\% C.L.\ by the total rates only, for $\tan^2\phi=0$
(solid curves) and $\tan^2\phi=0.06$ (dotted curves). For $\tan^2\phi=0.06$,
the SMA and LMA solutions are slightly shifted to lower values of
$\tan^2\omega$, while the LOW solution is shifted to higher values (including
the value $\omega=\pi/4$). 
}
\label{f13}
\end{figure}
\begin{figure}
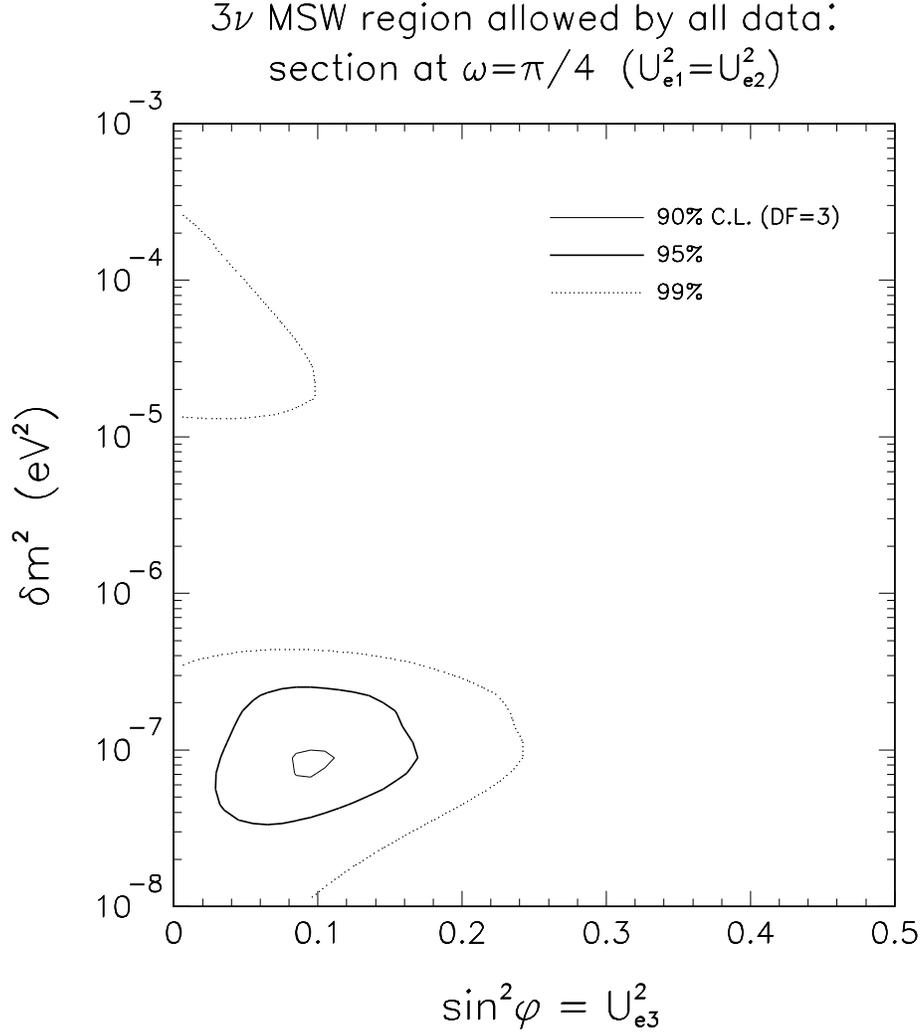

\caption{
Section of the allowed volume in the $3\nu$ parameter space in the plane
$(\delta m^2,\sin^2\phi)$, for the case of maximal $(\nu_1,\nu_2)$ mixing
$(\omega=\pi/4)$.  For $\sin^2\phi=0$, both the LMA and LOW solutions are
compatible with maximal mixing at 99\% C.L. For small values of $\sin^2\phi$,
the maximal mixing case favors the LOW solution.
}
\label{f14}
\end{figure}


\newcommand{\InsertFigure}[2]{\newpage\begin{center}\mbox{%
\epsfig{bbllx=1.4truecm,bblly=1.3truecm,bburx=19.5truecm,bbury=26.5truecm,%
height=21.truecm,figure=#1}}\end{center}\vspace*{-1.85truecm}%
\parbox[t]{\hsize}{\small\baselineskip=0.5truecm\hskip0.5truecm #2}}

\InsertFigure{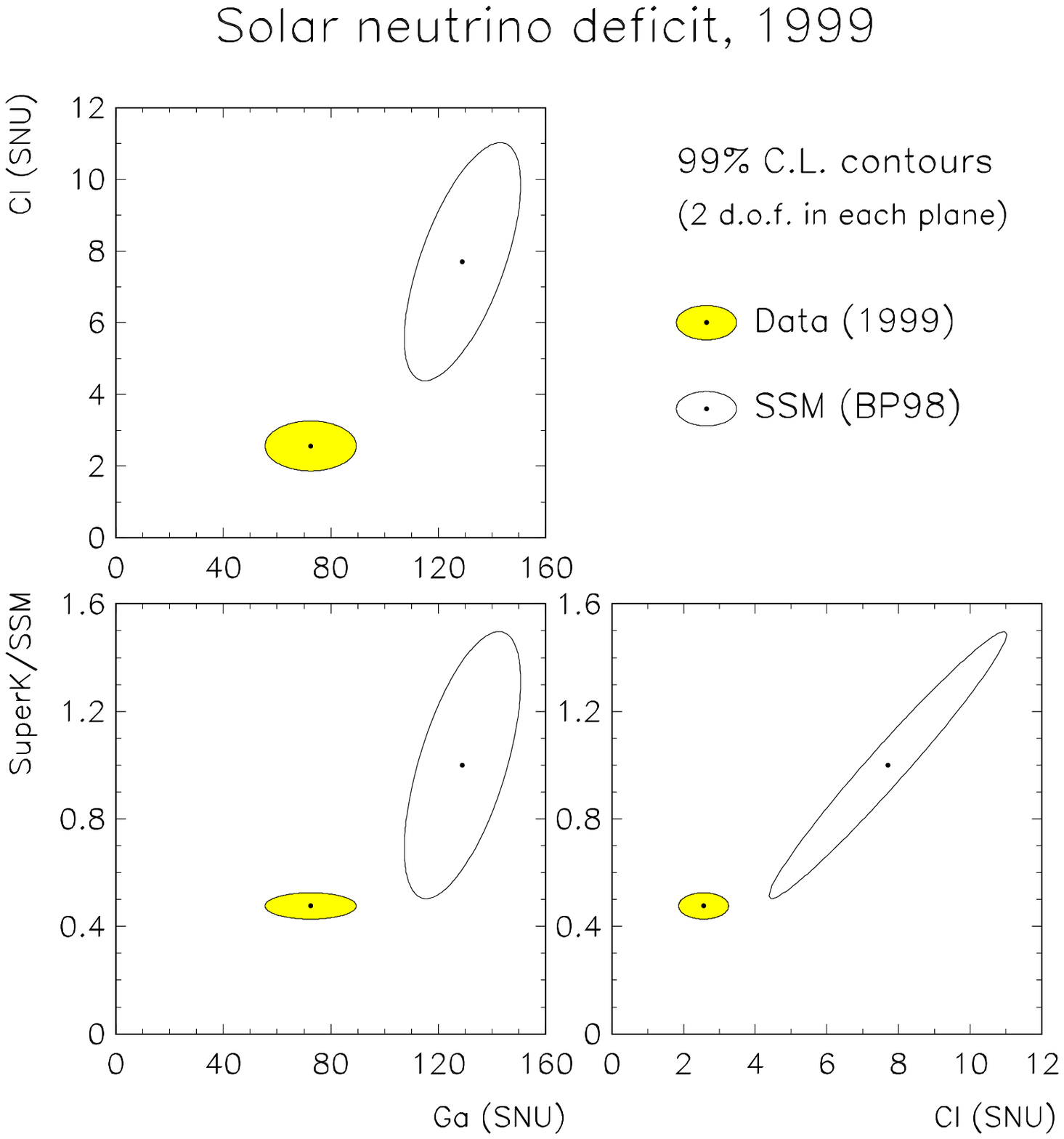}%
{FIG.~\protect\ref{f1}. 
The solar neutrino deficit, shown as a discrepancy between data and
expectations in the gallium (Ga), chlorine (Cl), and Super-Kamiokande total
event rates. In each plane, the error ellipses represent 99\% C.L.\ contours
for two degrees of freedom (i.e., $\Delta\chi^2=9.21$). The  projection of an
ellipse onto one of the axis gives approximately the $\pm3\sigma$ range for the
corresponding rate. Data and expectations refer to
Table~\protect\ref{tablerates}. The correlation of SSM errors is calculated as
in the Appendix.
}
\InsertFigure{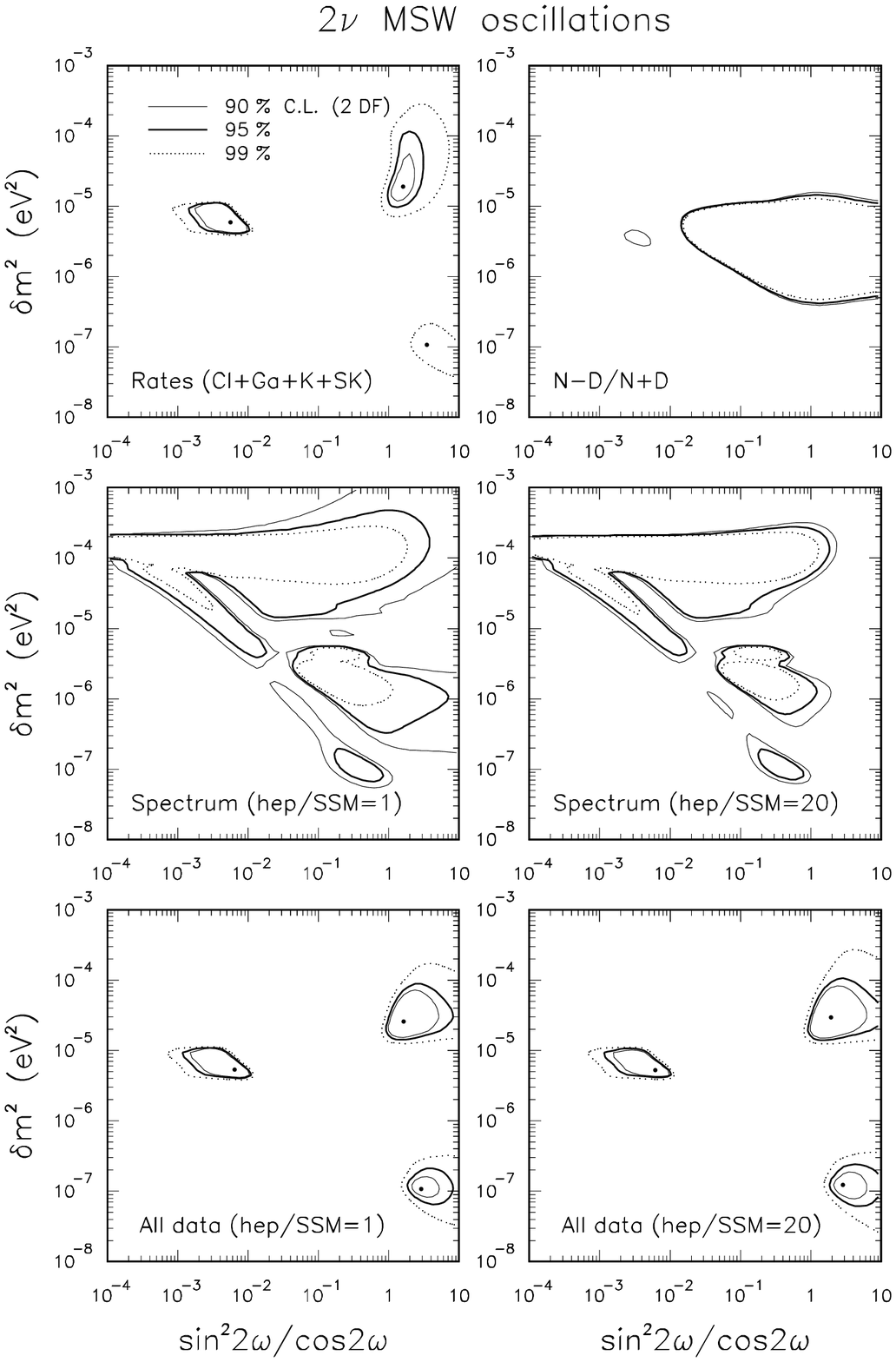}%
{FIG.~\protect\ref{f2}. 
Two-generation MSW solutions to the solar neutrino problem. The upper  four
panels correspond to the following separate fits to data subsets:  total rates
(Cl+Ga+K+SK); Super-Kamiokande night-day asymmetry $N-D/N+D$; Super-Kamiokande
electron energy spectrum with standard {\em hep\/} neutrino flux;
Super-Kamiokande spectrum with enhanced $(20\times)$ {\em hep\/} neutrino flux.
The two lower panels show the results of global fits to all data. The thin
solid, thick solid, and dashed curves correspond to $\chi^2-\chi^2_{\rm
min}=4.61$, 5.99, and 9.21. The positions of the local $\chi^2$ minima in fits
including the total rates are indicated by dots. See also
Tab.~\protect\ref{chisquare}.
}
\InsertFigure{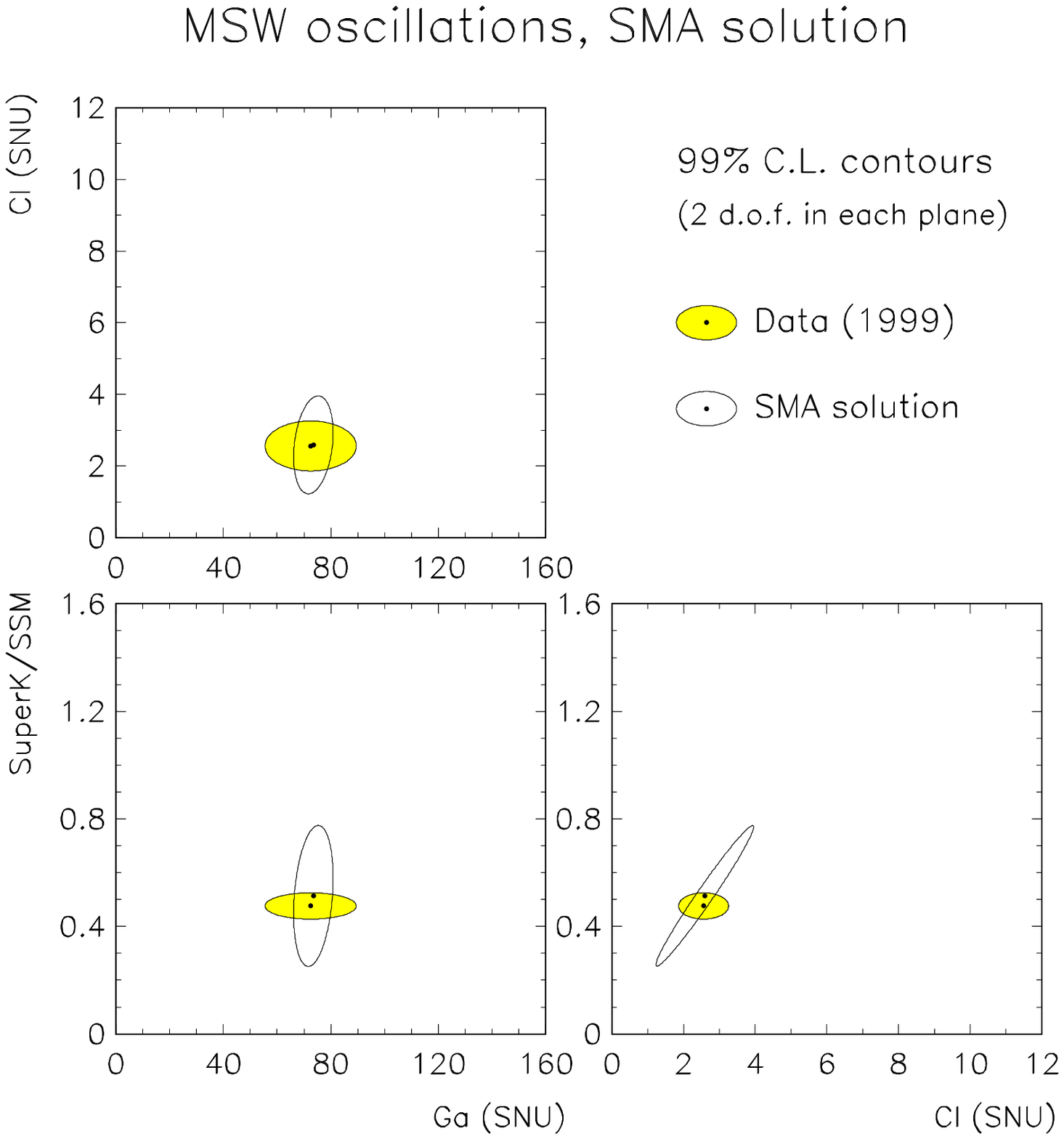}%
{FIG.~\protect\ref{f3}. 
The SMA solution at best fit (total rates only, first row of
Tab.~\protect\ref{chisquare}), compared with the experimental data, in the same
planes as in Fig.~\protect\ref{f1}. Note the excellent agreement between theory
and observations.
}
\InsertFigure{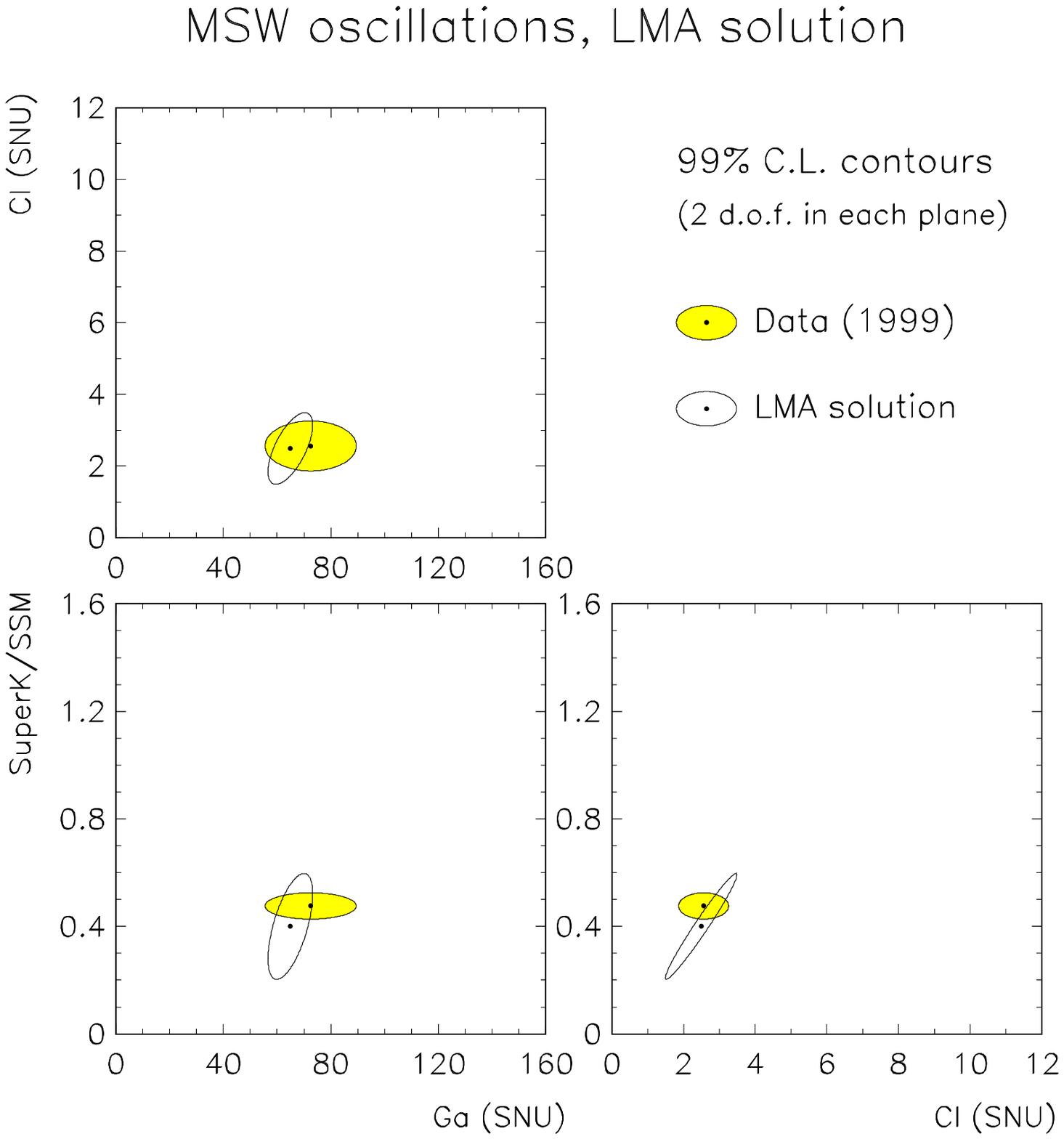}%
{FIG.~\protect\ref{f4}. 
As in Fig.~\protect\ref{f3}, but for the LMA solution at best fit  (total rates
only, second row of Tab.~\protect\ref{chisquare}).
}
\InsertFigure{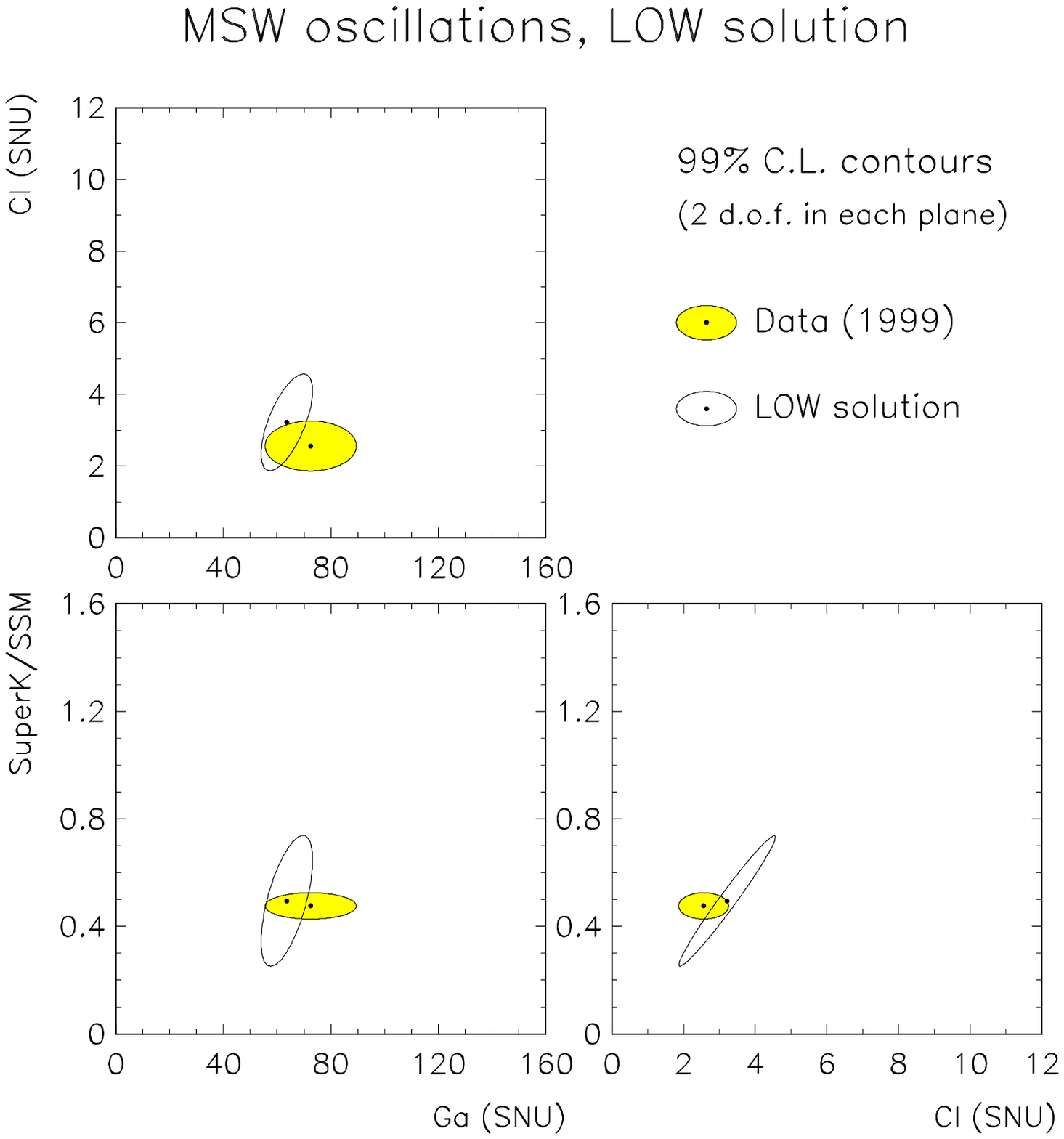}%
{FIG.~\protect\ref{f5}.
As in Fig.~\protect\ref{f3}, but for the LOW solution at best fit  (total rates
only, third row of Tab.~\protect\ref{chisquare}).
}
\InsertFigure{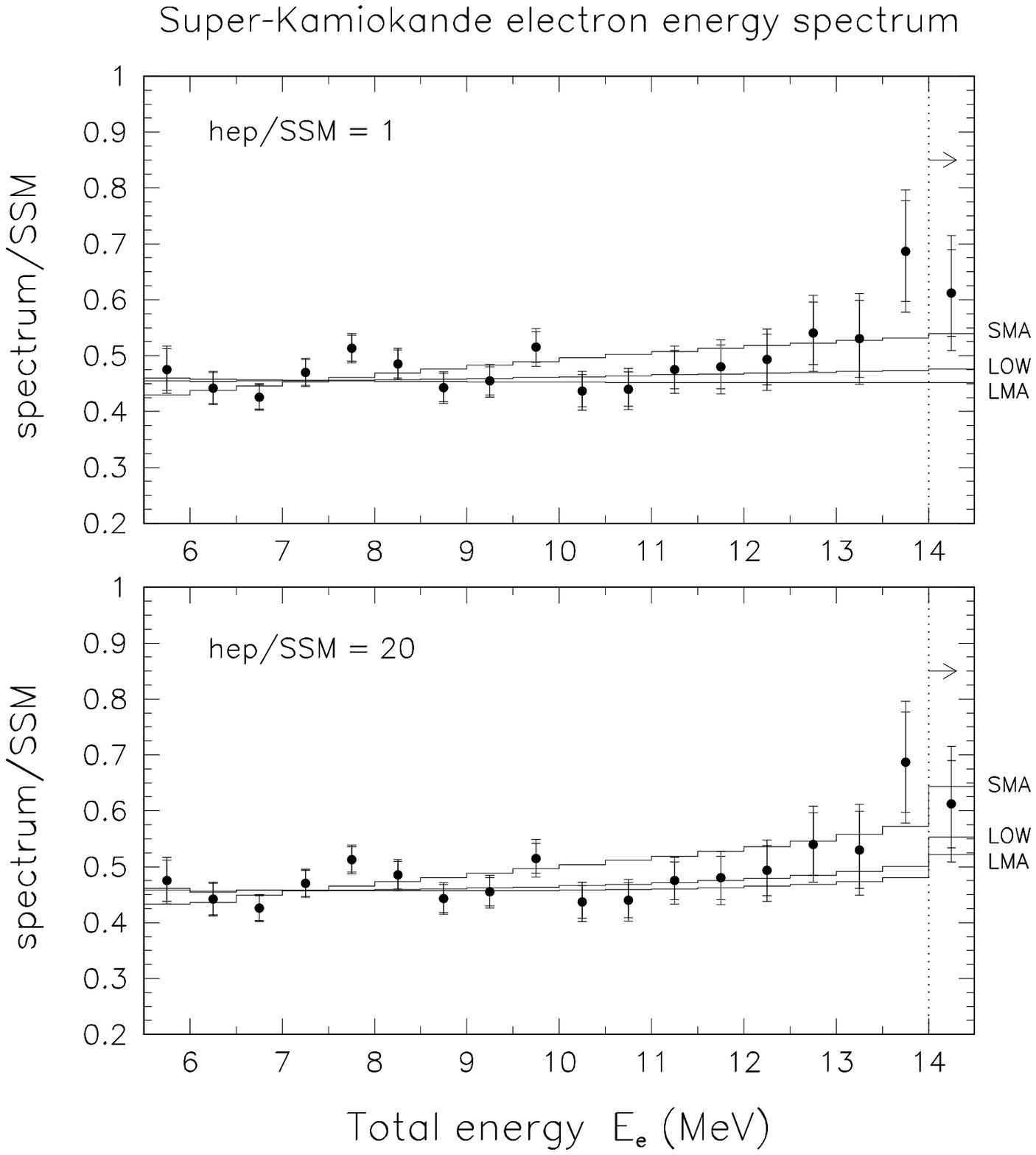}%
{FIG.~\protect\ref{f6}.
Comparison of data and predictions for the Super-Kamiokande spectrum shape. The
theoretical spectra normalization is taken free in the fit. The upper (lower)
panel corresponds to the case of standard  ($20\times$) {\em hep\/} flux. The
SMA, LMA, and LOW spectra are calculated in the global best fit points reported
in  Tab.~\protect\ref{chisquare} (middle rows for the upper panel and lower
rows for the lower panel). The SK data are reported from 
\protect\cite{To99,Su99,Na99}, and the error bars refer to statistical and
total experimental errors.
}
\InsertFigure{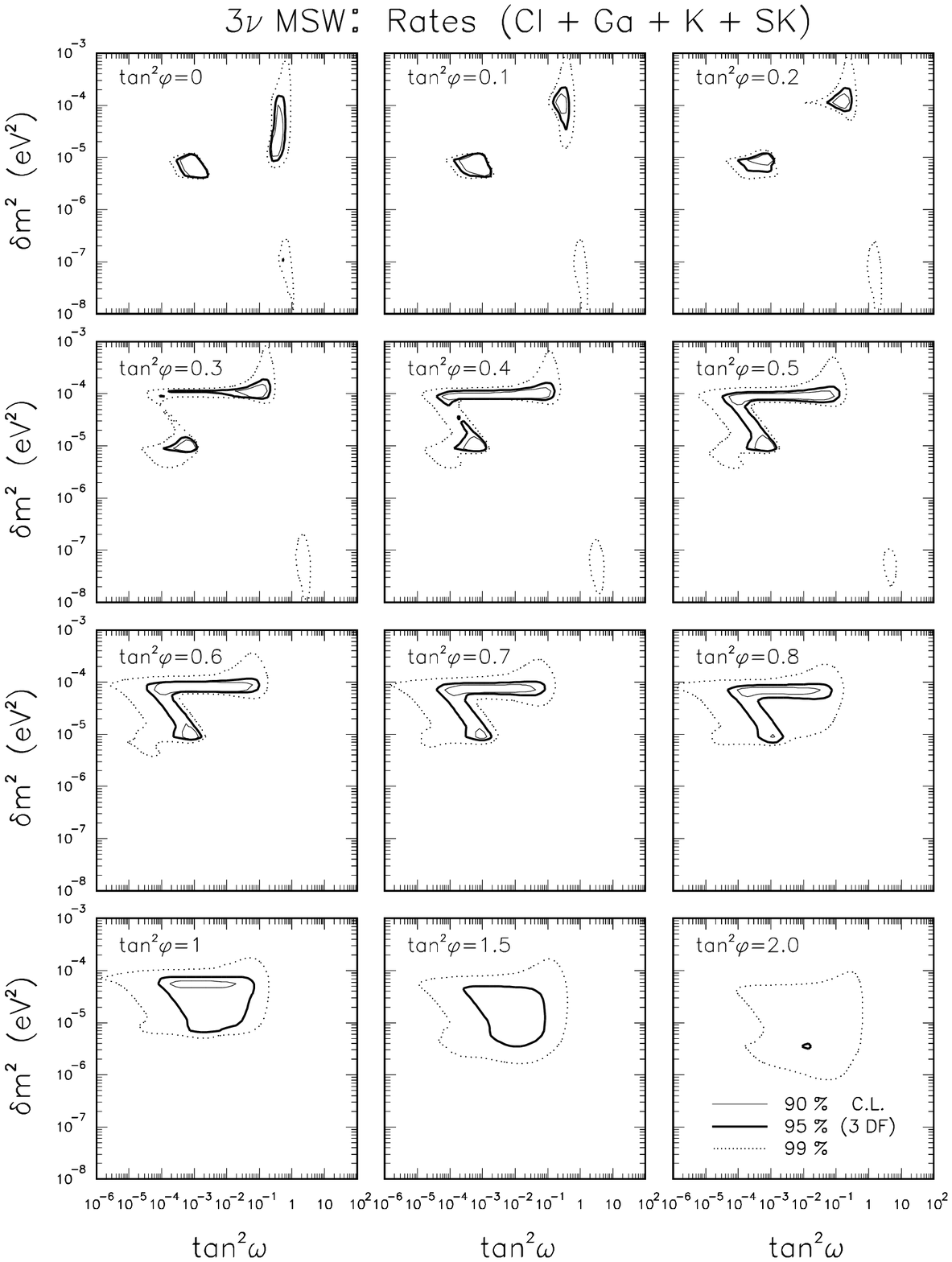}%
{FIG.~\protect\ref{f7}.
Three-flavor MSW oscillations: global fit to Cl+Ga+K+SK rates in the $(\delta
m^2,\tan^2\omega,\tan^2\phi)$ parameter space. The favored regions in each
panel correspond to sections of the volume  allowed at 90\%, 95\%, and 99\%
C.L.\  ($\chi^2-\chi^2_{\min}=6.25$, 7.82, and 11.36) for representative values
of $\tan^2\phi$.
}
\InsertFigure{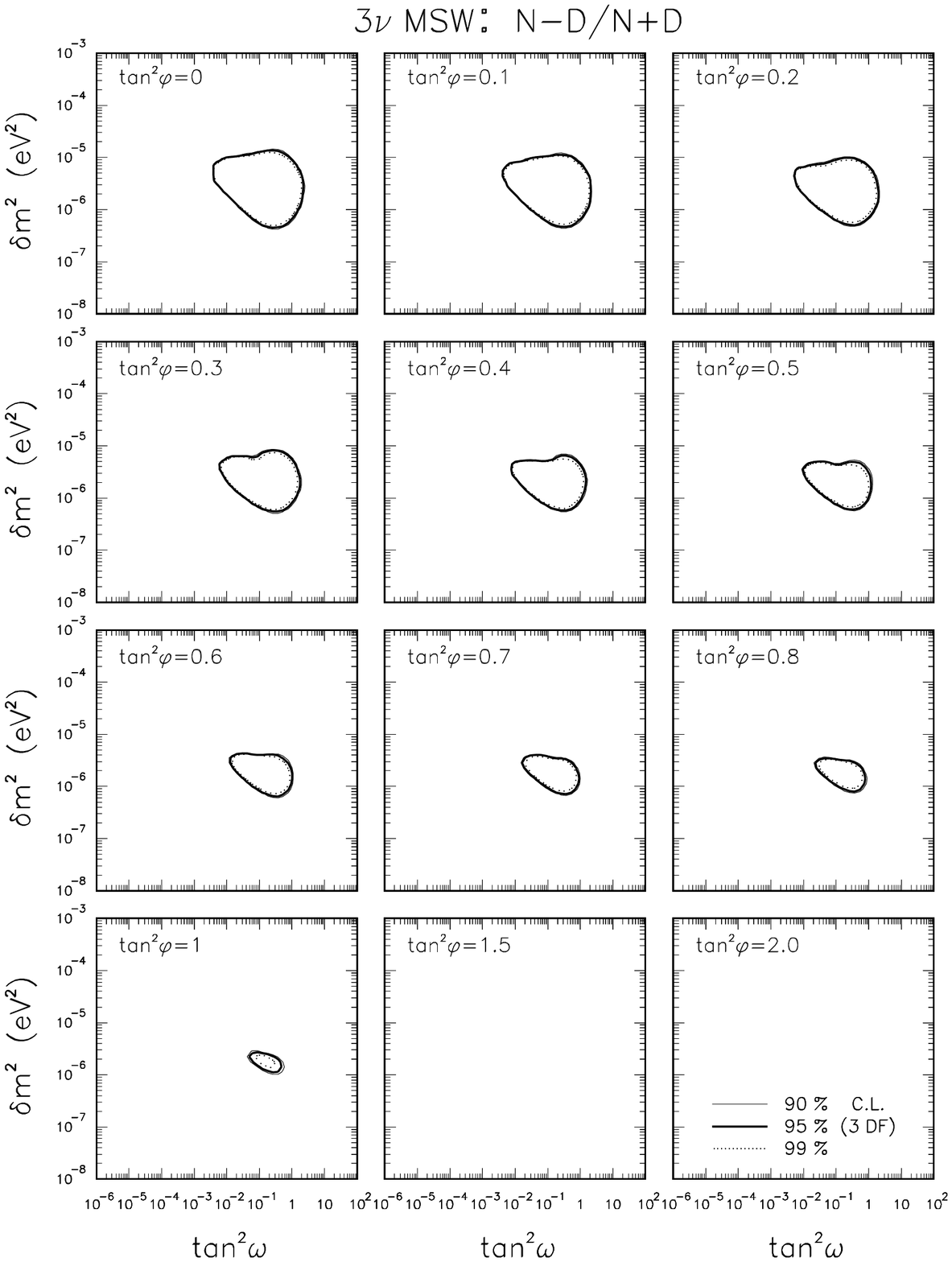}%
{FIG.~\protect\ref{f8}. 
As in Fig.~\protect\ref{f7}, but for the fit to the Super-Kamiokande night-day
asymmetry. The region inside the curves is excluded.
}
\InsertFigure{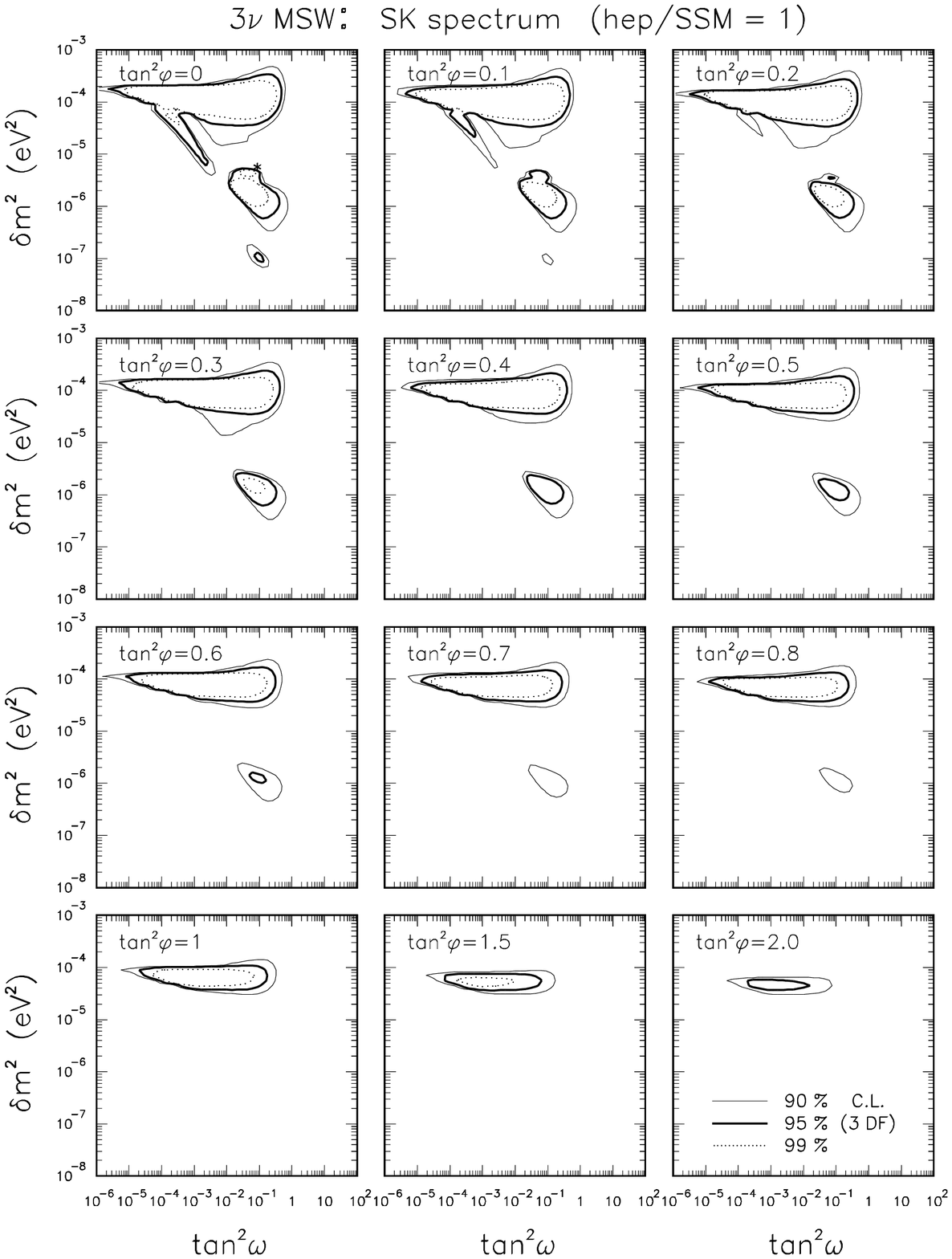}%
{FIG.~\protect\ref{f9}.
As in Fig.~\protect\ref{f7}, but for the fit to the Super-Kamiokande energy
spectrum. The regions inside the curves are excluded.
} 
\InsertFigure{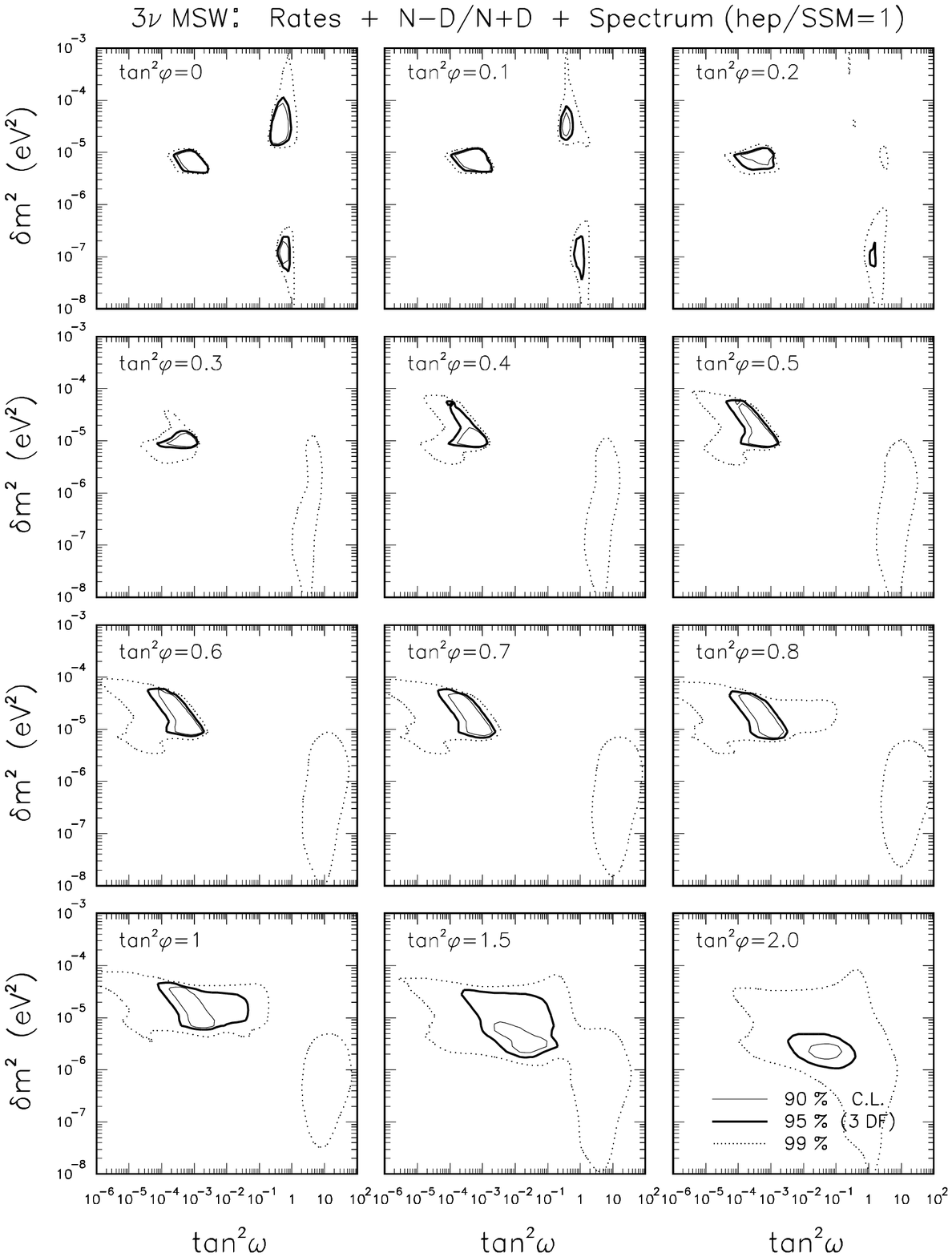}%
{FIG.~\protect\ref{f10}.
Results of the global three-flavor MSW fit to all data. Notice that, in the
first two panels, the 99\% C.L.\ contours  are compatible with maximal mixing
($\tan^2\omega=1$) for both the LOW and the LMA solutions.
}
\InsertFigure{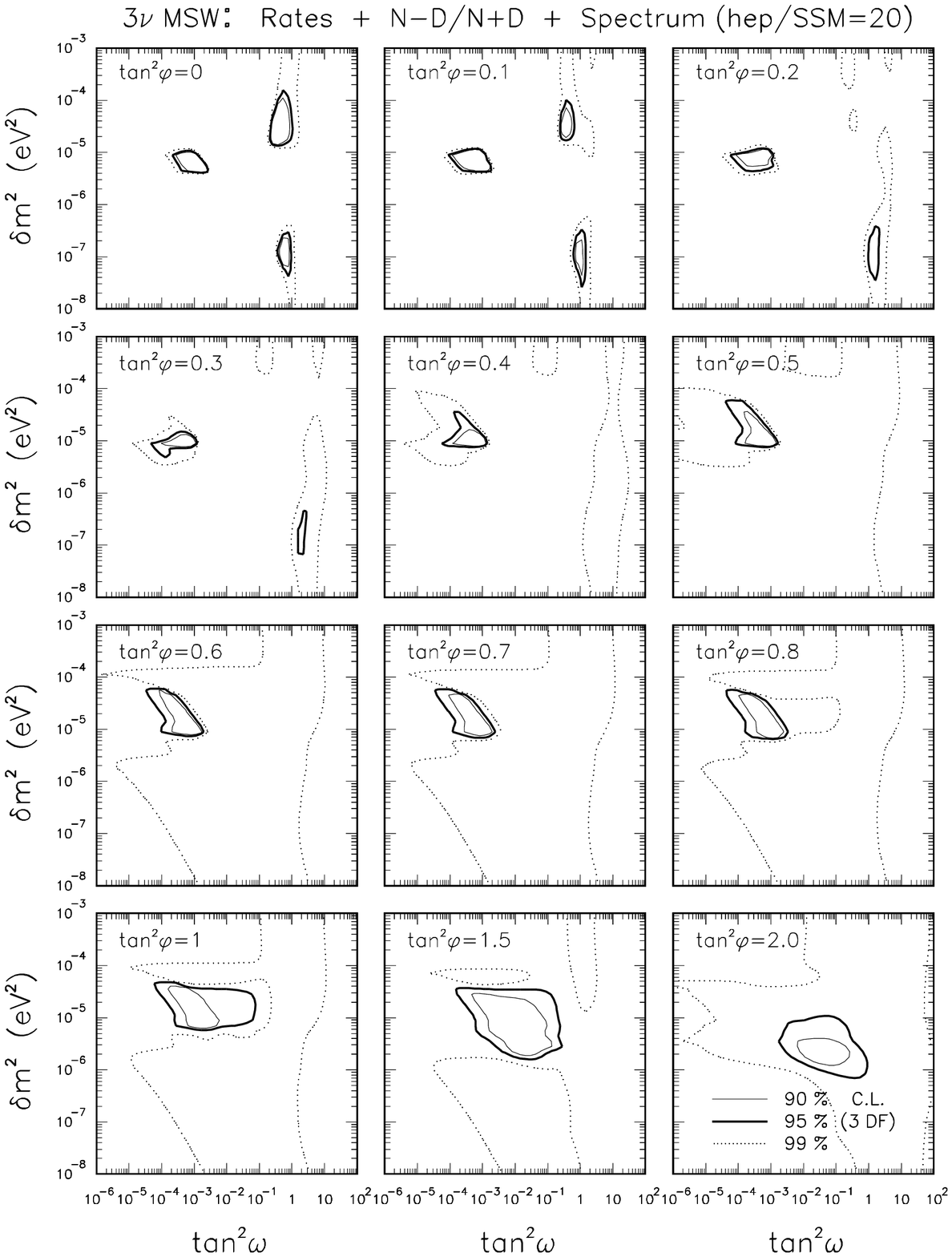}%
{FIG.~\protect\ref{f11}.
As in Fig.~\protect\ref{f10}, but for the case of enhanced ($20\times$) {\em
hep\/} flux. The allowed regions are slightly enlarged with respect to
Fig.~\protect\ref{f10}.
}
\InsertFigure{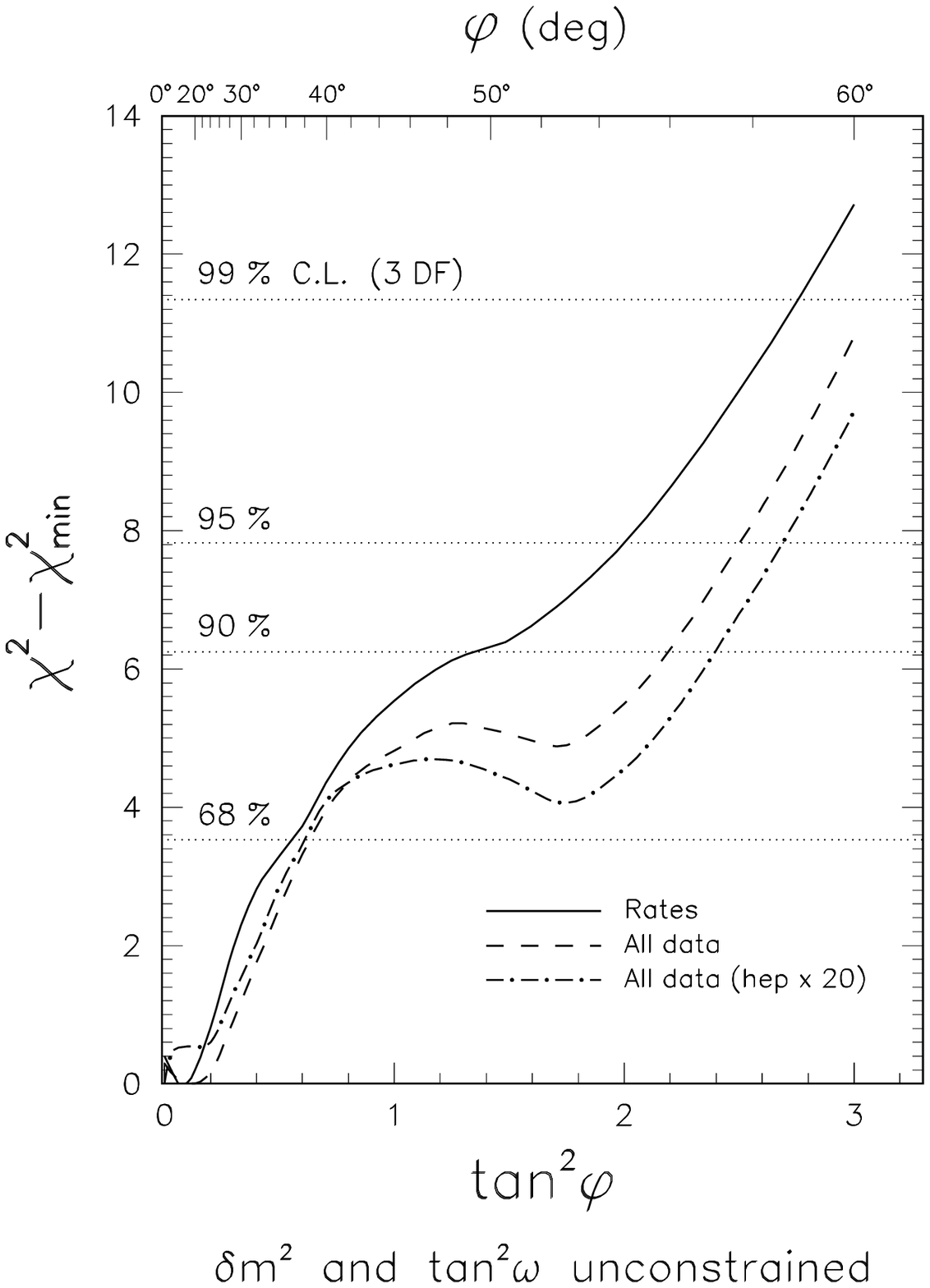}%
{FIG.~\protect\ref{f12}.
Values of $\Delta\chi^2$ as a function of $\phi$, for unconstrained $\delta
m^2$ and $\tan^2\omega$. At 95\% C.L., the upper limit on $\phi$  is in the
range $55^\circ$--$59^\circ$, depending on the data used in the fit and on the
value of the {\em hep\/} flux.
}
\InsertFigure{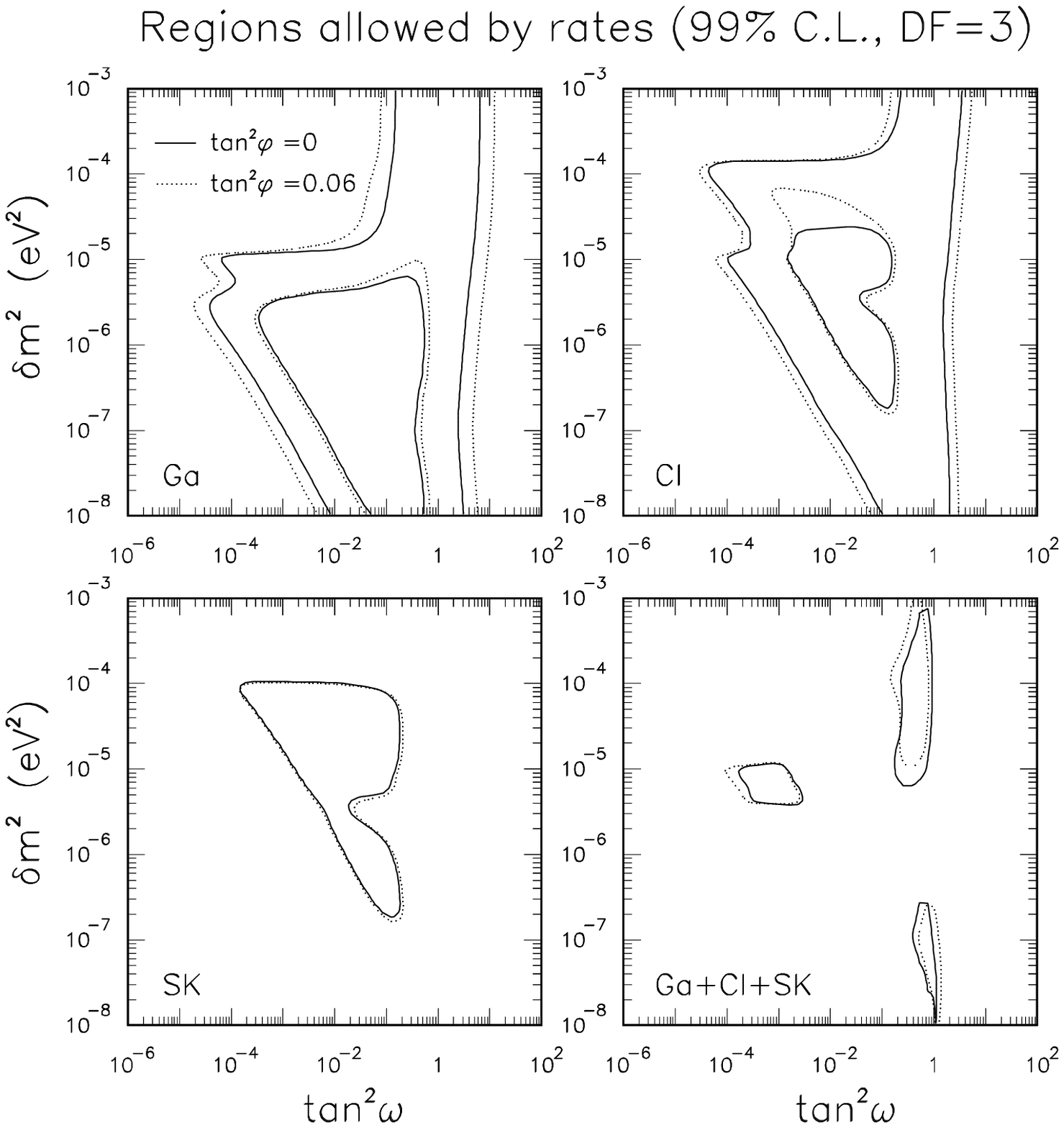}%
{FIG.~\protect\ref{f13}.
Regions allowed at 99\% C.L.\ by the total rates only, for $\tan^2\phi=0$
(solid curves) and $\tan^2\phi=0.06$ (dotted curves). For $\tan^2\phi=0.06$,
the SMA and LMA solutions are slightly shifted to lower values of
$\tan^2\omega$, while the LOW solution is shifted to higher values (including
the value $\omega=\pi/4$).
}
\InsertFigure{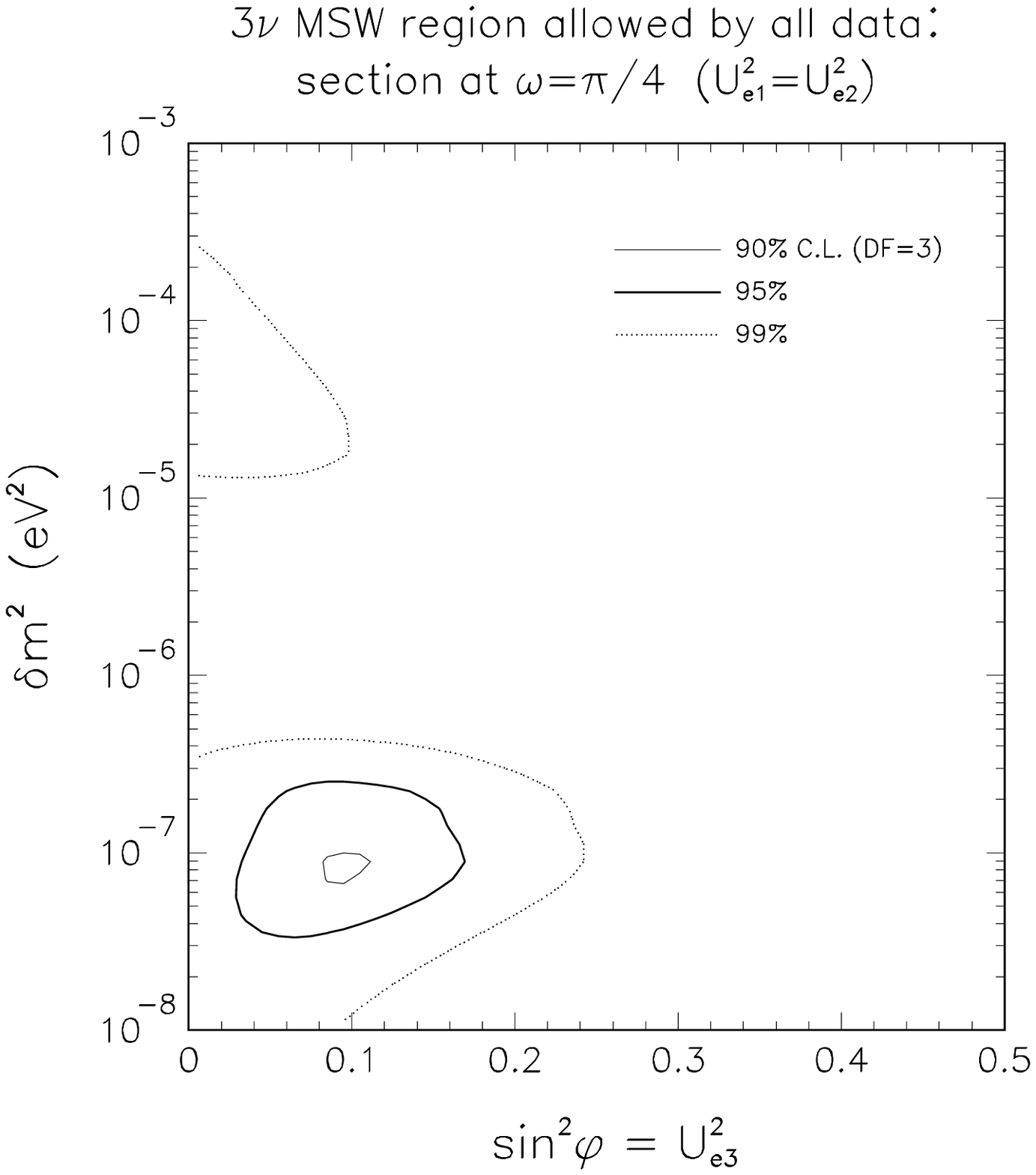}%
{FIG.~\protect\ref{f14}.
Section of the allowed volume in the $3\nu$ parameter space in the plane
$(\delta m^2,\sin^2\phi)$, for the case of maximal $(\nu_1,\nu_2)$ mixing
$(\omega=\pi/4)$.  For $\sin^2\phi=0$, both the LMA and LOW solutions are
compatible with maximal mixing at 99\% C.L. For small values of $\sin^2\phi$,
the maximal mixing case favors the LOW solution.
}

\end{document}